\documentclass[12pt]{article}

\usepackage{amsmath}
\usepackage{wrapfig}
\newcommand\pubnumber{SUHEP-010}
\newcommand\pubdate{\today}

\def\napoli{Physics Department\\
Syracuse University, Syracuse, New York, USA 13244-1130}
\def\support{\footnote{Work supported by the U.S. National Science Foundation.}}

\def\Title#1{\begin{center} {\Large #1 } \end{center}}
\def\Author#1{\begin{center}{ \sc #1} \end{center}}
\def\Address#1{\begin{center}{ \it #1} \end{center}}

\newcommand\pubblock{\rightline{\begin{tabular}{l} \pubnumber\\
         \pubdate  \end{tabular}}}
\newenvironment{Abstract}{\begin{quotation}  }{\end{quotation}}
\newenvironment{Presented}{\begin{quotation} \begin{center} 
             PRESENTED AT\end{center}\bigskip 
      \begin{center}\begin{large}}{\end{large}\end{center} \end{quotation}}
\def\Acknowledgements{\bigskip  \bigskip \begin{center} \begin{large}
             \bf ACKNOWLEDGEMENTS \end{large}\end{center}}

\usepackage{lineno}  
\usepackage{graphicx}  
\usepackage{xspace} 
\usepackage{color}
\usepackage{colortbl}
\usepackage{amsmath} 
\usepackage{ifthen} 
\usepackage{multirow}

\newboolean{pdflatex}
\setboolean{pdflatex}{true} 
\newboolean{articletitles}
\setboolean{articletitles}{true} 

\newboolean{uprightparticles}
\setboolean{uprightparticles}{false} 

\usepackage{amssymb}
\usepackage{amsfonts}
\usepackage{upgreek} 

\usepackage{hyperref}    

\usepackage[all]{hypcap} 
\usepackage{afterpage}

\ifthenelse{\boolean{uprightparticles}}%
{

 \def\Ppi         {\ensuremath{\uppi}\xspace}

 \def\Ppsi        {\ensuremath{\uppsi}\xspace}

 \def\PDelta      {\ensuremath{\Delta}\xspace}                 
 \def\PXi      {\ensuremath{\Xi}\xspace}                 
 \def\PLambda      {\ensuremath{\Lambda}\xspace}                 
 \def\PSigma      {\ensuremath{\Sigma}\xspace}                 
 \def\POmega      {\ensuremath{\Omega}\xspace}                 
 \def\PUpsilon      {\ensuremath{\Upsilon}\xspace}                 
 
 \def\PB      {\ensuremath{\mathrm{B}}\xspace}                 
                  
 \def\PD      {\ensuremath{\mathrm{D}}\xspace}

 \def\PJ      {\ensuremath{\mathrm{J}}\xspace}                 
 \def\PK      {\ensuremath{\mathrm{K}}\xspace}

 \def\Pb      {\ensuremath{\mathrm{b}}\xspace}                 
 \def\Pc      {\ensuremath{\mathrm{c}}\xspace}

 \def\Pi      {\ensuremath{\mathrm{i}}\xspace}

 \def\Ps      {\ensuremath{\mathrm{s}}\xspace}

}
{

 \def\Ppi         {\ensuremath{\pi}\xspace}

 \def\Ppsi        {\ensuremath{\psi}\xspace}                 
                  
 \mathchardef\PDelta="7101
 \mathchardef\PXi="7104
 \mathchardef\PLambda="7103
 \mathchardef\PSigma="7106
 \mathchardef\POmega="710A
 \mathchardef\PUpsilon="7107
                  
 \def\PB      {\ensuremath{B}\xspace}                 
                  
 \def\PD      {\ensuremath{D}\xspace}

 \def\PJ      {\ensuremath{J}\xspace}                 
 \def\PK      {\ensuremath{K}\xspace}

 \def\Pb      {\ensuremath{b}\xspace}                 
 \def\Pc      {\ensuremath{c}\xspace}

 \def\Pi      {\ensuremath{i}\xspace}

 \def\Ps      {\ensuremath{s}\xspace}

}

\makeatletter
\ifcase \@ptsize \relax
  \newcommand{\miniscule}{\@setfontsize\miniscule{4}{5}}
\or
  \newcommand{\miniscule}{\@setfontsize\miniscule{5}{6}}
\or
  \newcommand{\miniscule}{\@setfontsize\miniscule{5}{6}}
\fi
\makeatother

\DeclareRobustCommand{\optbar}[1]{\shortstack{{\miniscule (\rule[.5ex]{1.25em}{.18mm})}
  \\ [-.7ex] $#1$}}

\def\squark    {{\ensuremath{\Ps}}\xspace}

\def\cquark    {{\ensuremath{\Pc}}\xspace}

\def\bquark    {{\ensuremath{\Pb}}\xspace}

\def\pion   {{\ensuremath{\Ppi}}\xspace}

\def\pip    {{\ensuremath{\pion^+}}\xspace}
\def\pim    {{\ensuremath{\pion^-}}\xspace}

\def\kaon    {{\ensuremath{\PK}}\xspace}
  \def\Kbar    {{\kern 0.2em\overline{\kern -0.2em \PK}{}}\xspace}

\def\KorKbar    {\kern 0.18em\optbar{\kern -0.18em K}{}\xspace}

\def\Km      {{\ensuremath{\kaon^-}}\xspace}

\def\Kstarzb {{\ensuremath{\Kbar{}^{*0}}}\xspace}

  \def\Dbar    {{\kern 0.2em\overline{\kern -0.2em \PD}{}}\xspace}

\def\DorDbar    {\kern 0.18em\optbar{\kern -0.18em D}{}\xspace}

\def\B       {{\ensuremath{\PB}}\xspace}
\def\Bbar    {{\ensuremath{\kern 0.18em\overline{\kern -0.18em \PB}{}}}\xspace}

\def\BorBbar    {\kern 0.18em\optbar{\kern -0.18em B}{}\xspace}

\def\Bzb     {{\ensuremath{\Bbar{}^0}}\xspace}

\def\Bub     {{\ensuremath{\B^-}}\xspace}

\def\Bm      {{\ensuremath{\Bub}}\xspace}

\def\Bs      {{\ensuremath{\B^0_\squark}}\xspace}
\def\Bsb     {{\ensuremath{\Bbar{}^0_\squark}}\xspace}
\def\Bdb     {{\ensuremath{\Bbar{}^0}}\xspace}

\def\jpsi     {{\ensuremath{{\PJ\mskip -3mu/\mskip -2mu\Ppsi\mskip 2mu}}}\xspace}

  \def\Y#1S{\ensuremath{\PUpsilon{(#1S)}}\xspace}

\def\Xires       {{\ensuremath{\PXi}}\xspace}

\def\Lz          {{\ensuremath{\PLambda}}\xspace}
\def\Lbar        {{\ensuremath{\kern 0.1em\overline{\kern -0.1em\PLambda}}}\xspace}
\def\LorLbar    {\kern 0.18em\optbar{\kern -0.18em \PLambda}{}\xspace}

\def\Lb      {{\ensuremath{\Lz^0_\bquark}}\xspace}

\def\Lc      {{\ensuremath{\Lz^+_\cquark}}\xspace}

\def\Xibz    {{\ensuremath{\Xires^0_\bquark}}\xspace}

\def\to                 {\ensuremath{\rightarrow}\xspace}

\def\CP                {{\ensuremath{C\!P}}\xspace}

\def\AT#1     {\ensuremath{A_{\mathrm{T}}^{#1}}\xspace}           

\def\C#1      {\ensuremath{\mathcal{C}_{#1}}\xspace}                       
\def\Cp#1     {\ensuremath{\mathcal{C}_{#1}^{'}}\xspace}                    
\def\Ceff#1   {\ensuremath{\mathcal{C}_{#1}^{\mathrm{(eff)}}}\xspace}        
\def\Cpeff#1  {\ensuremath{\mathcal{C}_{#1}^{'\mathrm{(eff)}}}\xspace}       
\def\Ope#1    {\ensuremath{\mathcal{O}_{#1}}\xspace}                       
\def\Opep#1   {\ensuremath{\mathcal{O}_{#1}^{'}}\xspace}                    

\newcommand{\bra}[1]{\ensuremath{\langle #1|}}             
\newcommand{\ket}[1]{\ensuremath{|#1\rangle}}              

\newcommand{\tev}{\ifthenelse{\boolean{inbibliography}}{\ensuremath{~T\kern -0.05em eV}\xspace}{\ensuremath{\mathrm{\,Te\kern -0.1em V}}}\xspace}
\newcommand{\gev}{\ensuremath{\mathrm{\,Ge\kern -0.1em V}}\xspace}
\newcommand{\mev}{\ensuremath{\mathrm{\,Me\kern -0.1em V}}\xspace}
\newcommand{\kev}{\ensuremath{\mathrm{\,ke\kern -0.1em V}}\xspace}
\newcommand{\ev}{\ensuremath{\mathrm{\,e\kern -0.1em V}}\xspace}
\newcommand{\gevc}{\ensuremath{{\mathrm{\,Ge\kern -0.1em V\!/}c}}\xspace}
\newcommand{\mevc}{\ensuremath{{\mathrm{\,Me\kern -0.1em V\!/}c}}\xspace}
\newcommand{\gevcc}{\ensuremath{{\mathrm{\,Ge\kern -0.1em V\!/}c^2}}\xspace}
\newcommand{\gevgevcccc}{\ensuremath{{\mathrm{\,Ge\kern -0.1em V^2\!/}c^4}}\xspace}
\newcommand{\mevcc}{\ensuremath{{\mathrm{\,Me\kern -0.1em V\!/}c^2}}\xspace}

\def\invfb   {\ensuremath{\mbox{\,fb}^{-1}}\xspace}

\def\gsim{{~\raise.15em\hbox{$>$}\kern-.85em
          \lower.35em\hbox{$\sim$}~}\xspace}
\def\lsim{{~\raise.15em\hbox{$<$}\kern-.85em
          \lower.35em\hbox{$\sim$}~}\xspace}

\def\tell1  {TELL1\xspace}
\def\ukl1   {UKL1\xspace}

 \renewcommand{\tev}{\ensuremath{\mathrm{\,Te\kern -0.1em V}}\xspace}
\renewcommand{\gev}{\ensuremath{\mathrm{\,Ge\kern -0.1em V}}\xspace}
\renewcommand{\mev}{\ensuremath{\mathrm{\,Me\kern -0.1em V}}\xspace}
\renewcommand{\kev}{\ensuremath{\mathrm{\,ke\kern -0.1em V}}\xspace}
\renewcommand{\ev}{\ensuremath{\mathrm{\,e\kern -0.1em V}}\xspace}
\renewcommand{\gevc}{\ensuremath{{\mathrm{\,Ge\kern -0.1em V\!/}c}}\xspace}
\renewcommand{\mevc}{\ensuremath{{\mathrm{\,Me\kern -0.1em V\!/}c}}\xspace}
\renewcommand{\gevcc}{\ensuremath{{\mathrm{\,Ge\kern -0.1em V\!/}c^2}}\xspace}
\renewcommand{\gevgevcccc}{\ensuremath{{\mathrm{\,Ge\kern -0.1em V^2\!/}c^4}}\xspace}
\renewcommand{\mevcc}{\ensuremath{{\mathrm{\,Me\kern -0.1em V\!/}c^2}}\xspace}
\usepackage{accents}

\usepackage{cite}
\usepackage{mciteplus}

\begin{document}
\begin{titlepage}
\pubblock

\vfill
\Title{Lifetimes of some $b$-flavored hadrons}
\vfill
\Author{ Sheldon Stone\support}
\Address{\napoli}
\vfill
\begin{Abstract}
Recent measurements of lifetimes of some $b$-flavored hadrons are presented and interpreted in the context of theoretical models, especially the Heavy Quark Expansion. Decay widths and decay width differences  in the $\Bs-\Bsb$ system are discussed from studies of decays into the final states $\jpsi K^+K^-$, $\jpsi \pi^+\pi^-$, $D_s^+D_s^-$, $K^+K^-$, and $D_s^{\pm}\pi^{\mp}$. Lifetime measurements of the baryons $\Lb$, $\Xi_b^-$, $\Xi_b^0$, and $\Omega_b^-$ are also shown.
\end{Abstract}
\vfill
\begin{Presented}
2014 Flavor Physics and CP Violation (FPCP-2014)\\
Marseille, France,  May 26--30, 2014
\end{Presented}
\vfill
\end{titlepage}
\def\thefootnote{\fnsymbol{footnote}}
\setcounter{footnote}{0}

\section{Introduction}

Lifetimes of elementary particles contain important information about the interactions that govern their decays.  Theoretical models worthy of consideration must predict lifetimes, or ratios of lifetimes, accurately.
Decay time distributions are  basically exponential but in neutral meson decays  can be modified by both mixing and \CP violation. However, this is a crude way to learn about \CP violation.
One model, called the Heavy Quark Expansion, HQE, is used to determine $|V_{cb}|$ and $|V_{ub}|$. It  can be tested by using its predictions for relative $b$-hadron lifetimes \cite{Lenz:2014jha}.
At lowest order the $b$-quark decay governs the lifetime, except for $B_c$ decays that are not covered here, but higher order corrections are important.

Let us first consider the decay of \Bsb mesons. Since we sum over \Bs and \Bsb decays to measure lifetimes, the decay time distribution into a given final state $f$ is given by \cite{Dunietz:2000cr,*Dunietz:1986vi}
\begin{eqnarray}
\Gamma[f,t]&=&\Gamma\left(\Bs(t)\to f\right)+\Gamma\left(\Bsb(t)\to f\right)\nonumber\\
&=&{\cal{N}}_f\left[e^{-\Gamma_Lt}\left|\bra{f}{B_L}\rangle\right|^2+e^{-\Gamma_Ht}\left|\bra{f}{B_H}\rangle\right|^2\right]\nonumber\\
&=&{\cal{N}}_f\left|A_f\right|^2\left[1+|\lambda_f|^2\right]e^{-\Gamma_st}\left\{\cosh\frac{\Delta\Gamma_st}{2}+{\cal{A}}_{\Delta\Gamma}\sinh\frac{\Delta\Gamma_st}{2}\right\},
\end{eqnarray}
where $\ket{B_L}$ and $\ket{B_H}$ are the mass eigenstates with corresponding decay widths $\Gamma_L$ and $\Gamma_H$, so that $\Delta\Gamma_s=\Gamma_L-\Gamma_H$, and $\Gamma_s=(\Gamma_L+\Gamma_H)/2$. The amplitude for $\Bs\to f$ is given by $A_f$, while for $\Bsb\to f$ the amplitude is described by $\overline{A}_f$. The parameters $p$ and $q$ relate the mass eigenstates to the flavor eigenstates:  $\ket{B_L}=p\ket{\Bs}+q\ket{\Bsb}$ and $\ket{B_H}=p\ket{\Bs}-q\ket{\Bsb}$.
 In addition, ${\cal{A}}_{\Delta\Gamma}\equiv -2{\rm Re}(\lambda_f)/\left(1+|\lambda_f|^2\right)$, where $\lambda_f=\frac{q}{p}\frac{\overline{A}_f}{A_f}$. The decay time shape is not exponential and depends on the specific decay mode. To second order in $\Delta\Gamma_st$
\begin{equation}
\label{eq:KK}
\Gamma[f,t] \propto e^{-\Gamma_s t}\left[1+\frac{1}{2}\left(\frac{\Delta\Gamma_s}{2}t\right)^2+{\cal{A}}_{\Delta\Gamma}\left(\frac{\Delta\Gamma_s}{2}t\right)\right].
\end{equation}
In this paper we work in units where $\hbar=c=1$, so the lifetime $\tau=1/\Gamma$.

While the above equations equally can be applied to $\Bzb\equiv \overline{B}_d$ decays, $\Delta\Gamma_d/\Gamma_d$ has been
measured as 0.015$\pm$0.018 by the $e^+e^-$ B-factories \cite{PDG} , so the decay time distribution  can be treated as purely exponential given the derived limit $\Delta\Gamma_d<0.032$\,ps$^{-1}$ @ 95\% confidence level (CL),  consistent with the theoretical prediction in the Standard Model (SM) of $2\times 10^{-3}$\,ps$^{-1}$ \cite{Lenz:2011ti}.  This measurement should be pursued as the sensitivity is not close to the theoretical prediction and physics beyond the SM may well appear \cite{Gershon:2010wx,*Bobeth:2014rda}.

For \Bsb decays, $\Delta\Gamma_s$ is not small and ${\cal{A}}_{\Delta\Gamma}$ depends on the decay mode, mainly through $\overline{A}_f/A_f$ as $1-|q/p|$ has been measured as being small \cite{HFAG}.
For ``flavor specific" \Bsb decay modes, where $\Bsb\to\overline{f}$ and $\Bs\to f$, the decay is the sum of two exponentials that when fit with a single exponential can be approximated as \cite{Hartkorn:1999ga}
\begin{equation}
\label{eq:fs}
\Gamma_s\approx\Gamma_{\rm flavor~specific}\frac{1-\left(\frac{\Delta\Gamma_s}{2\Gamma_s}\right)^2}{1+\left(\frac{\Delta\Gamma_s}{2\Gamma_s}\right)^2}.
\end{equation}


I distinguish between two different methods of measuring lifetimes here. The usual method is to measure lifetimes absolutely, by fitting distributions of decay times to appropriate exponential or modified distributions after determining acceptances. For example,  the \Bzb lifetime is measured as 1.519$\pm$0.005\,ps, from many sources \cite{HFAG}, the most precise being the $e^+e^-$ B-factory experiments.

Another strategy is to measure the ratio of lifetimes of two different $B$ species. This usually done with the same decay topologies, for example $\Bsb\to\jpsi f_0(980)$, $f_0(980)\to \pi^+\pi^-$ compared with $\Bzb\to\jpsi \overline{K}^{*0}(980)$, $\overline{K}^{*0}(980)\to K^-\pi^+$  \cite{Aaij:2012nta} and then use the precisely determined \Bzb lifetime to extract the \Bsb lifetime. The advantage of this method is that many systematic uncertainties associated with the decay time acceptance are eliminated.

\section{\boldmath Widths and width differences in \Bsb decays}
To find a value for $\Gamma_s$ I will use absolute decay time measurements in the $\jpsi K^+K^-$ and $\jpsi\pi^+\pi^-$ decay modes, since all three observables governing the decay width $\Gamma_s$, $\Delta\Gamma_s$, and ${\cal{A}}_{\Delta\Gamma}$ are determined simultaneously. The $K^+K^-$ mode is restricted to the low mass region, dominated by the $\phi$ with a small component of S-wave \cite{Aaij:2013orb}  that was predicted \cite{Stone:2008ak}, while the $\pi^+\pi^-$ mode  has a large $f_0(980)$ contribution \cite{Aaij:2014emv} that was also predicted \cite{Stone:2008ak}. The suitability of using the  $\jpsi\pi^+\pi^-$ mode was questioned \cite{Fleischer:2011au} because the $f_0(980)$ was postulated as being a tetraquark state. However,  a recent full Dalitz analysis of the final state shows that this is not the case \cite{Aaij:2014siy}, based on the model of \cite{Stone:2013eaa}. 

\begin{figure}[t!]
\centering
\includegraphics[height=4in]{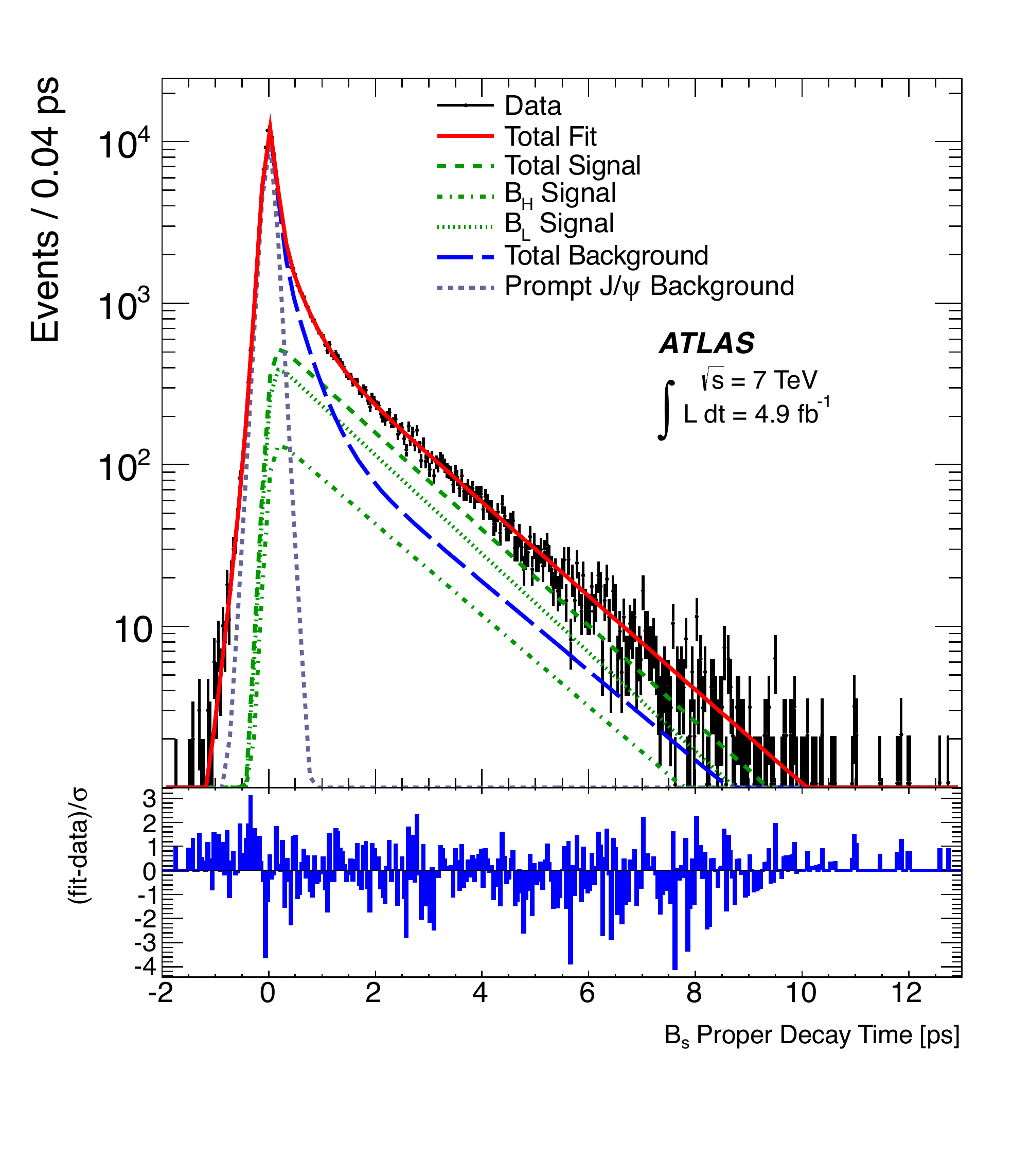}
\vspace{-1cm}
\caption{Proper decay time fit projection for $\Bsb\to\jpsi\phi$. The pull distribution at the bottom shows
the difference between the data and fit value normalized to the data
uncertainty.}
\label{ATLAS-Bslife}
\end{figure}

Since  the decay time distribution is modified by the presence of  \CP violation, even if \Bs and \Bsb decays are summed over (see Ref.~\cite{Aaij:2014emv} for a discussion), only measurements that determine simultaneously the width, the width difference and the \CP violating phase are considered. These measurements are listed in Table~\ref{tab:Gammas}. This approach differs from the current HFAG scheme of averaging all measurements including those in \CP eigenstate modes, by estimating ${\cal{A}}_{\Delta\Gamma}$ from SM theory. An example of the decay time distributions and signal fits is given in Fig.~\ref{ATLAS-Bslife} from the ATLAS collaboration \cite{Aad:2012kba}.

\begin{table}[htb]
\begin{center}
\begin{tabular}{l|cccc}  
Exp. &  $\int{\cal{L}}$ (fb$^{-1}$) & $\Gamma_s$ (ps$^{-1}$)   & $\Delta\Gamma_s$ (ps$^{-1}$) \\ \hline
ATLAS \cite{Aad:2012kba} & 4.9 &$0.6770\pm 0.0070\pm 0.0040$& $0.053\pm 0.021\pm 0.010$ \\
CDF \cite{Aaltonen:2012ie} & 9.6 &$0.6545\pm 0.0081\pm 0.0039$& $0.068\pm 0.026\pm 0.009$  \\
D0  \cite{Abazov:2011ry} & 8.0 &$0.6930\pm 0.0182$~~~~~~~~~~~~~& $0.163\pm 0.065$~~~~~~~~~~~  \\
LHCb \cite{Aaij:2013oba}& 1.0 &$0.6610\pm 0.0040\pm 0.0060$&$0.106\pm 0.011\pm 0.007$ \\\hline 
Average& & $0.6662\pm0.0045$ & $0.106\pm 0.013$\\ \hline
\end{tabular}
\caption{Measurements of $\Gamma_s$ and $\Delta\Gamma_s$ in $\Bsb$ plus \Bs $\to\jpsi\phi$ and $\jpsi\pi^+\pi^-$ decays.}
\label{tab:Gammas}
\end{center}
\end{table}

\begin{figure}[htb]
\centering
\includegraphics[height=3in]{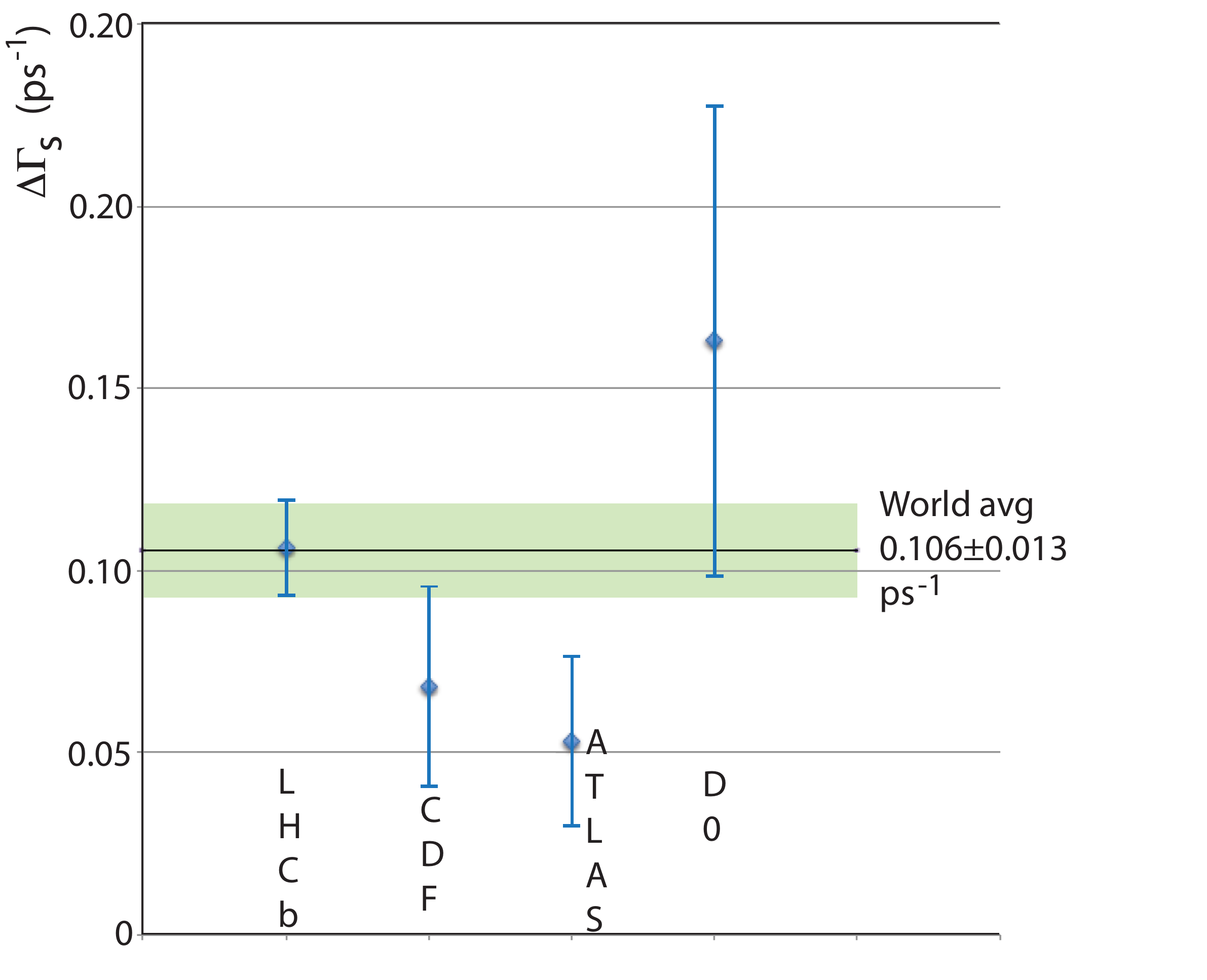}
\caption{Measurements of $\Delta\Gamma_s$ using the $\jpsi \phi$ and $\jpsi\pi^+\pi^-$ final states.}
\label{DGs}
\end{figure}

The average values are $\Gamma_s=0.6662\pm0.0045$\,ps$^{-1}$, and $\Delta\Gamma_s=0.106\pm 0.013$\,ps$^{-1}$, leading to 
a ratio $\Delta\Gamma_s/\Gamma_s=0.16\pm 0.02$. 
While $\Gamma_s$ is known with an impressive precision of 0.7\%, $\Delta\Gamma_s$ is much less well determined with the measurements widely scattered. Figure~\ref{DGs} shows the different measurements and my average, which is consistent with theoretical predictions \cite{Lenz:2006hd,Lenz:2011ti}.
It is also important for the interpretation of these results that \CP violation in these modes is measured to be small. The recent LHCb results, summarized by Artuso at this meeting \cite{Artuso-FPCP} average to $70\pm 55$\,mrad, sufficiently small to be ignored when only the \CP violation due to mixing is in play.

To get a better idea on the size of $\Delta\Gamma_s$ it is useful to look at decays into specific eigenstates. The final state $\jpsi f_0(980)$ is \CP odd, corresponding to the $\ket{B_H}$ state, while $D_s^+D_s^-$ and $K^+K^-$ are \CP even corresponding to the $\ket{B_L}$ state. The $\jpsi f_0(980)$ and $D_s^+D_s^-$ modes are expected to be dominated by tree level decays and thus direct \CP violation resulting from interference with Penguin diagrams is expected to be small. On the other hand the $K^+K^-$ mode is expected to be affected. 

\begin{figure}[b!]
\begin{center}
    \includegraphics[width=0.495\textwidth]{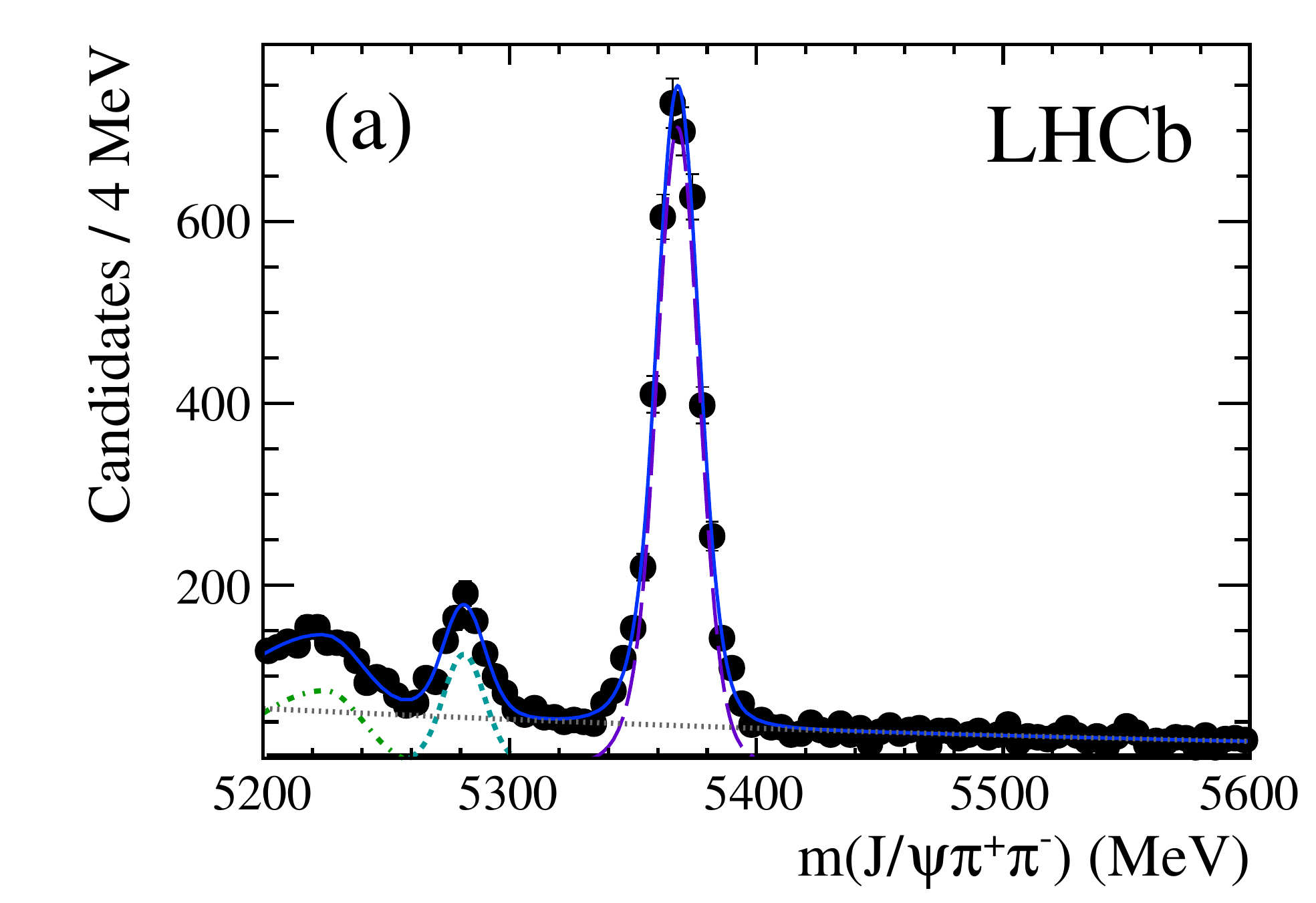}
    \includegraphics[width=0.495\textwidth]{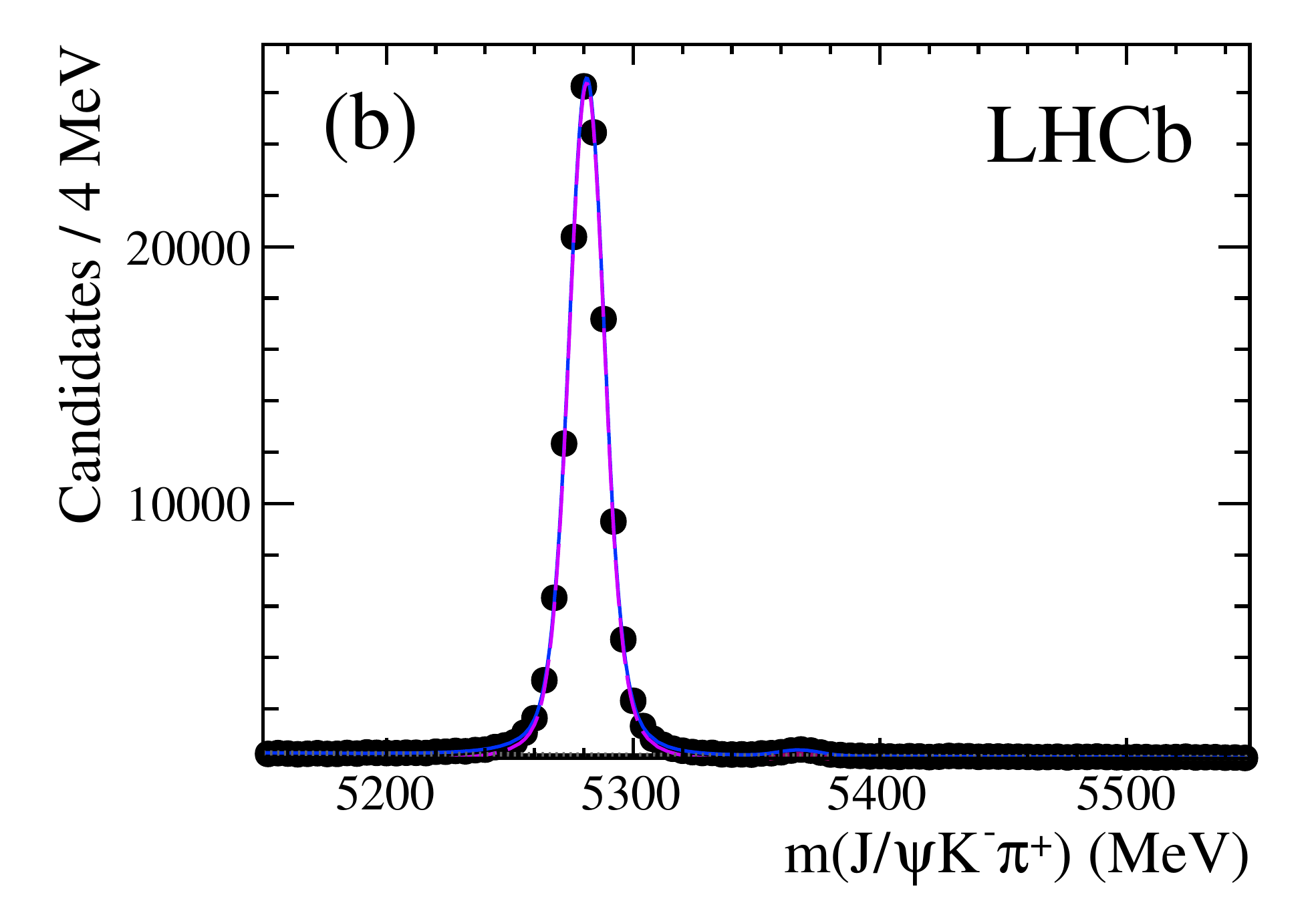}
    \vspace{-8mm}
	\caption{Invariant mass distributions of selected (a) $\jpsi\pi^+\pi^-$ and (b) $\jpsi\Km\pim$ candidates. The dipion candidates have been selected to be in the $f_(980)$ mass region and the $K^-\pi^+$ candidates selected to be near the $\overline{K}^{*0}(890)$. 
}
	\label{f:masspsi}
\end{center}
\end{figure}

\begin{figure}[t]
\begin{center}
	    \includegraphics[width=0.6\textwidth]{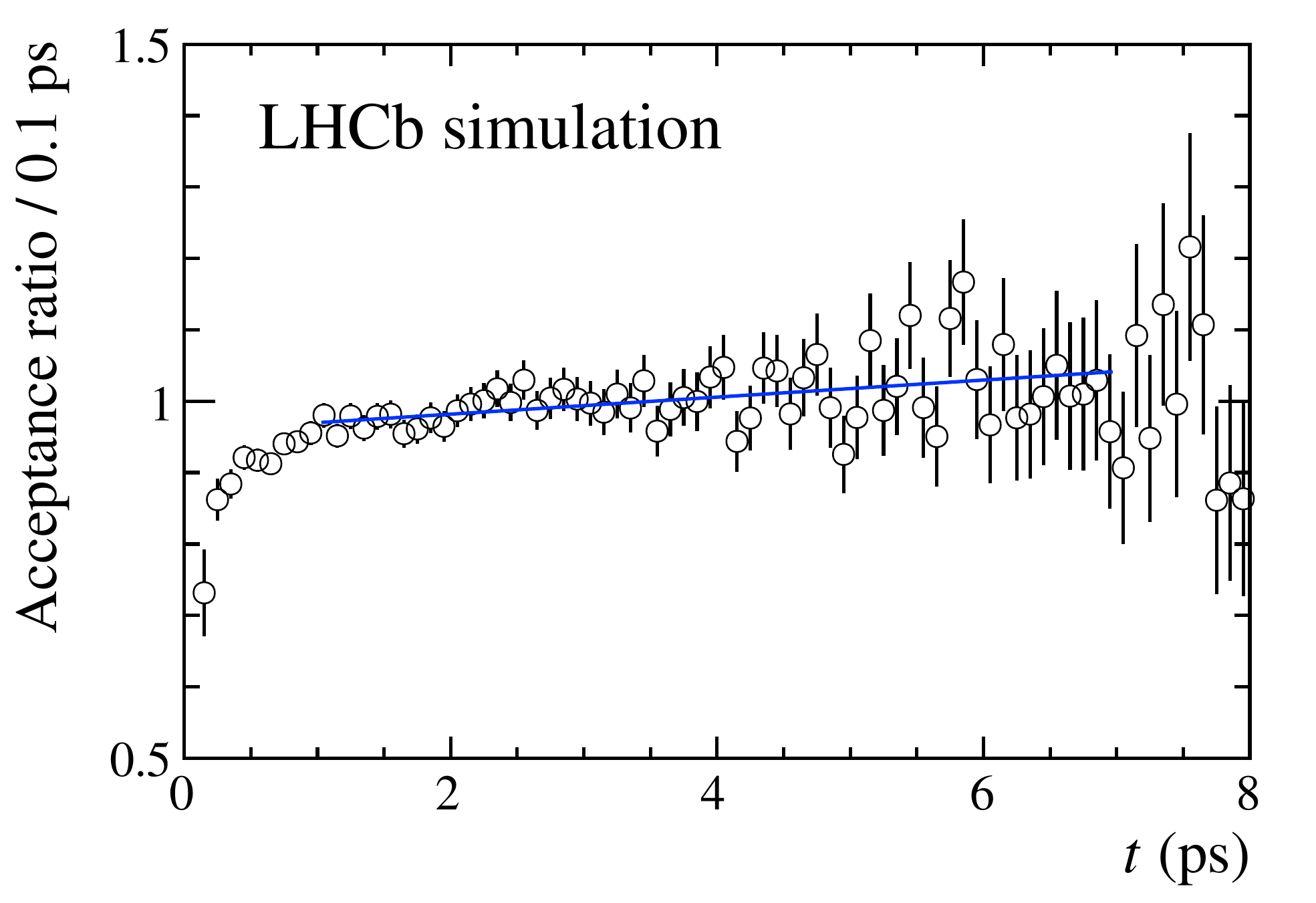}
	    \vspace{-2mm}
	\caption{Ratio of decay time acceptances obtained from simulation between $\Bsb\to\jpsi f_0(980)$ and $\Bzb\to\jpsi K^{*0}(890)$ decays.The solid (blue) line shows the result of a linear fit.}
		\label{jpsif0-Kstar-accept}
\end{center}
\end{figure}

The relative measurement of the lifetime in the mode $\Bsb\to\jpsi f_0(980)$ with respect to $\Bzb\to\jpsi \overline{K}^{*0}(890)$ was made by LHCb \cite{Aaij:2012nta}. The invariant mass spectra of the signal decay products is shown in Fig.~\ref{f:masspsi} using 1\,fb$^{-1}$ of data.  There are 4040$\pm$75 $\jpsi f_0(980)$ and 131,900$\pm$400 $\jpsi \overline{K}^{*0}(890)$ signal events, respectively.  Since the decay time acceptance is not quite the same for both modes, due to the different kinematics in the decay, simulation is used to determine the relative acceptance. This is shown in Fig.~\ref{jpsif0-Kstar-accept}. The lifetime is computed as $1.700\pm 0.040 \pm 0.26$\,ps. This result should not be included in an average with the previous one for $\Gamma_s$, as the data have already been used as part of the $\jpsi\pi^+\pi^-$ final state in the LHCb measurement of $\Gamma_s$. It is useful, however, to view this component separately.

\begin{figure}[t!]
\vspace{-2mm}
\centering
\includegraphics[width=0.49\textwidth]{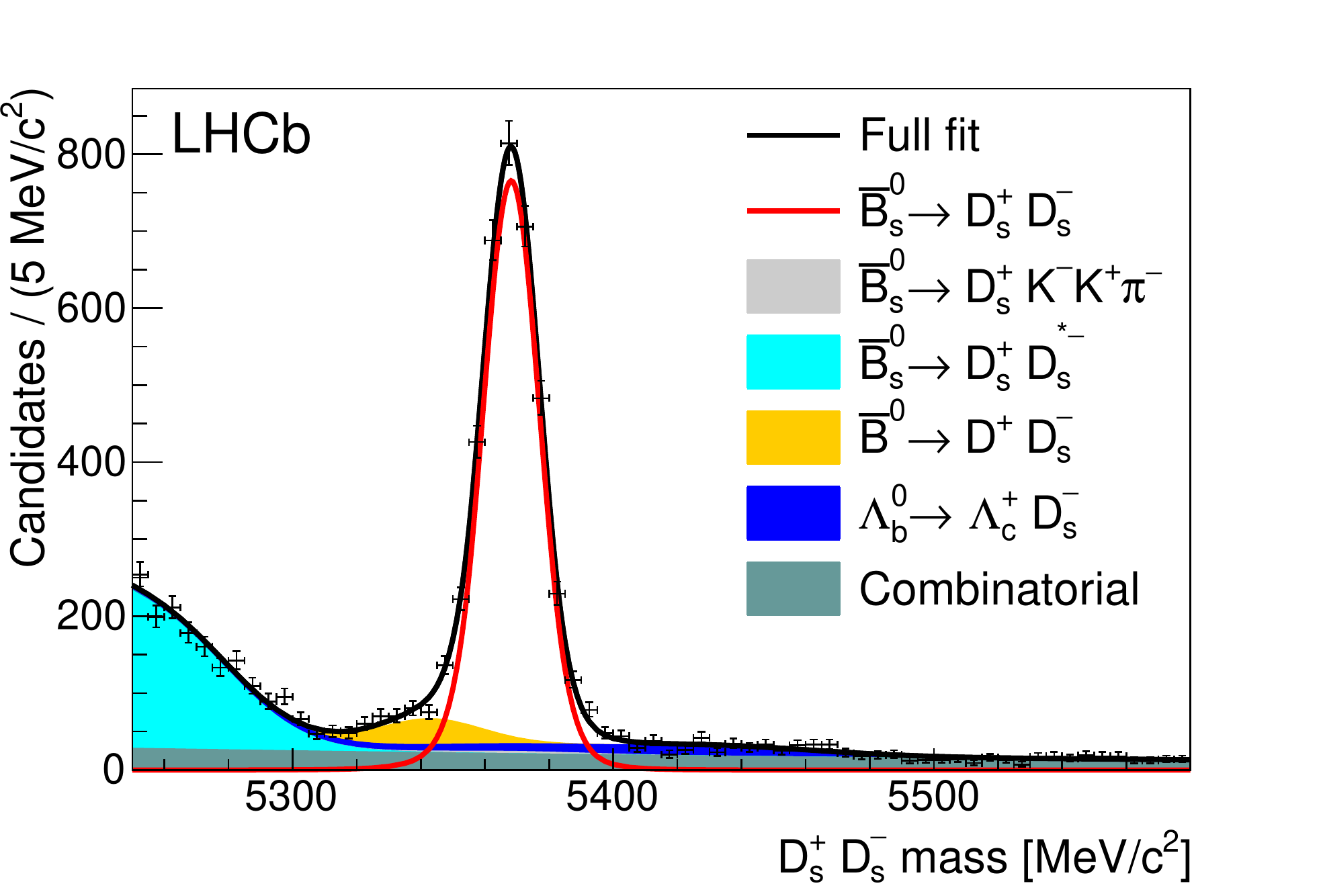}
\includegraphics[width=0.49\textwidth]{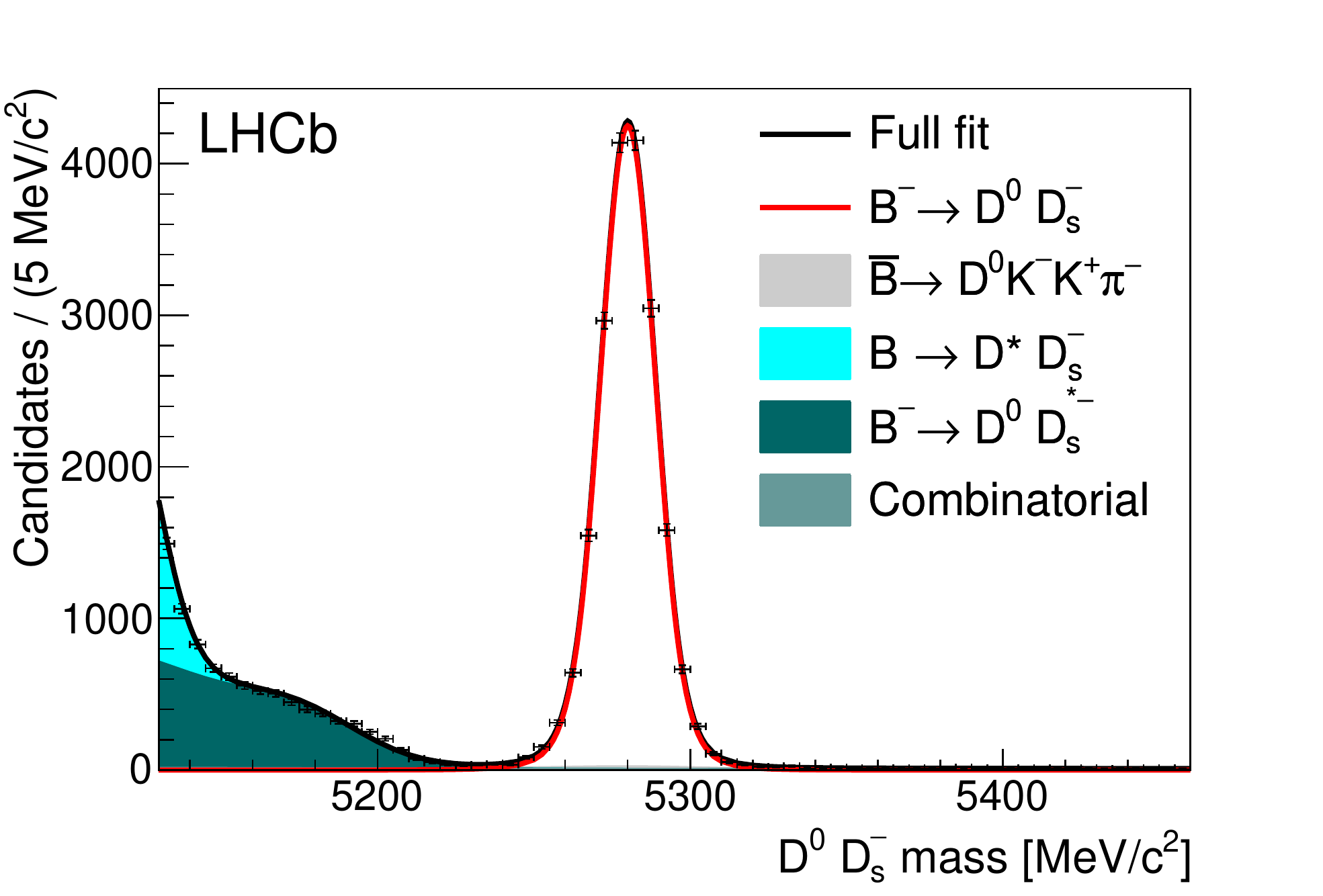}
\vspace{-3mm}
\caption{Mass distributions and fits to the full data sample for (left) $\Bsb\to D_s^+D_s^-$ and (right) $\Bm\to D^0D_s^-$
candidates. The points are the data and the curves and shaded regions show
the fit components. }
\label{fig:fitFullSample}
\end{figure}

Let us turn to the \CP-even $\Bsb\to D_s^+D_s^-$ decay mode that has been analyzed by LHCb \cite{Aaij:2013bvd}.
The invariant mass distributions for the signal, and the normalization mode
are shown in Fig.~\ref{fig:fitFullSample}, along
with the results of binned maximum likelihood fits for the various components.
In total, there are  3499\,$\pm$\,65 $\Bsb\to D_s^+D_s^-$ and 19,432\,$\pm$\,140 $\Bm\to D^0D_s^-$ decays. The lifetime is measured to be $1.379\pm0.026\pm0.017$\,ps.

\begin{figure}[htb]
\centering
\includegraphics[width=0.7\textwidth]{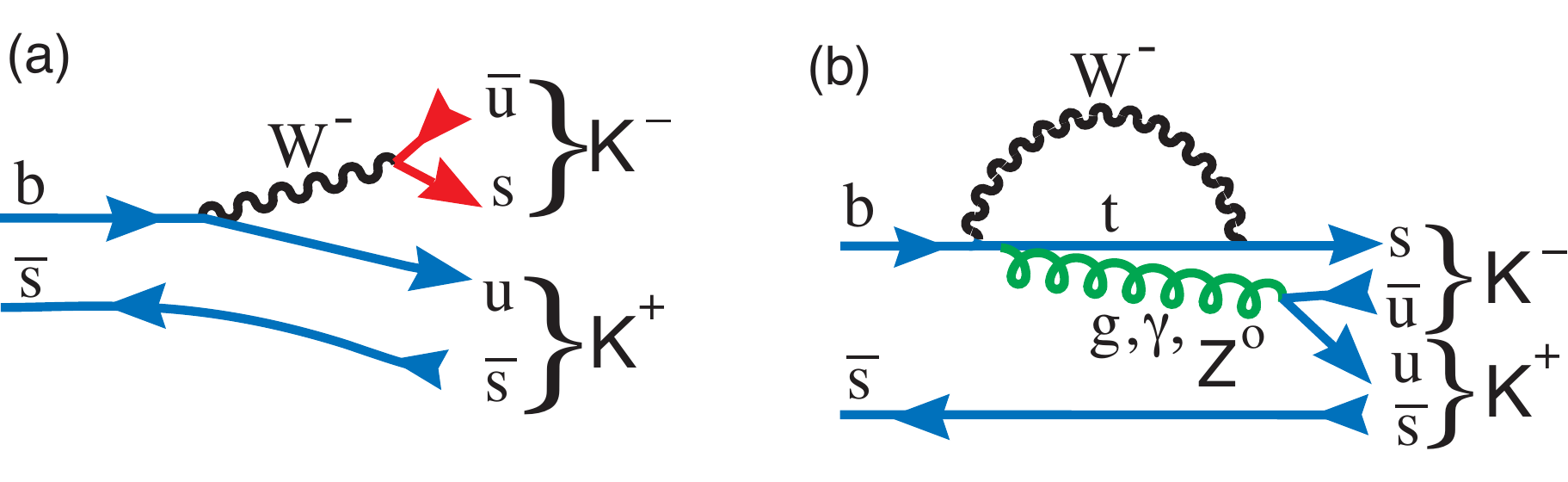}
\vspace{-3mm}
\caption{Feynman diagrams for $\Bsb\to K^+K^-$ decay, (a) Tree and (b) Penguin. }
\label{KK}
\end{figure}
Next we return to an absolute lifetime measurement done in the $\Bsb\to K^+K^-$ mode by LHCb. This mode can have both Tree and Penguin diagram contributions as illustrated in Fig.~\ref{KK}. 

\begin{figure}[t!]
\centering
\includegraphics[width=1.0\textwidth]{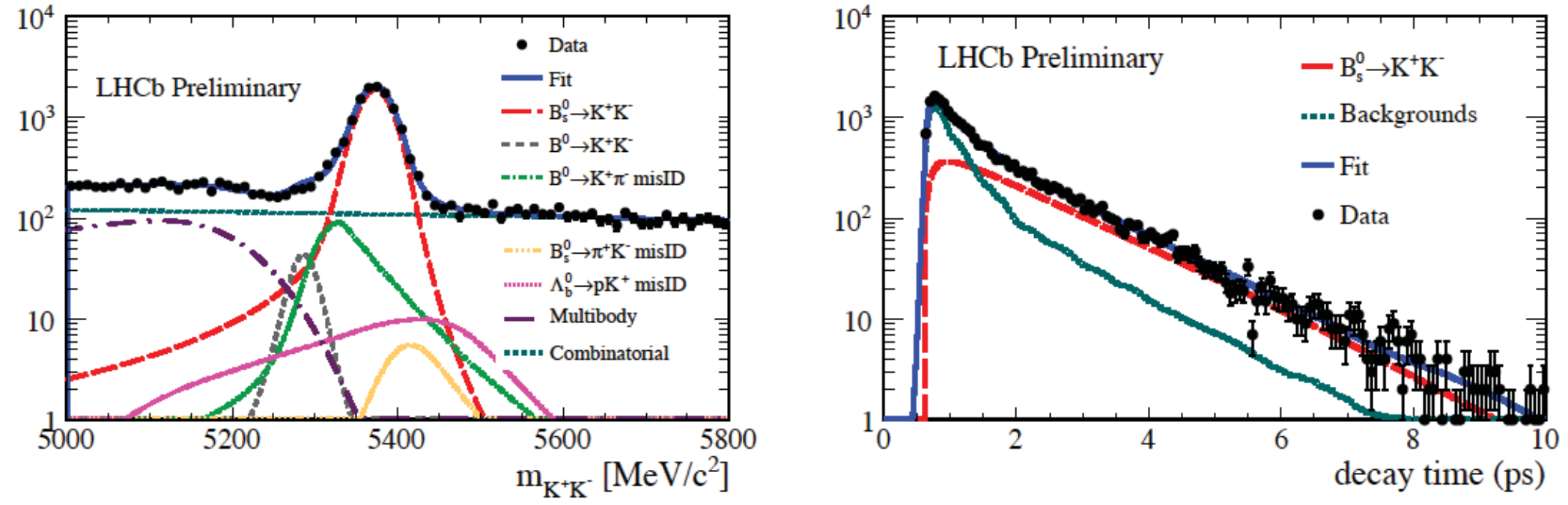}
\vspace{-6mm}
\caption{The invariant $K^+K^-$ mass spectrum (left) and the decay time distribution for  $\Bsb\to K^+K^-$ decay (right) along with backgrounds and the fit. }
\label{KK-life}
\end{figure}

The $K^+K^-$ invariant mass distribution for this mode is shown in Fig.~\ref{KK-life}(left). There is clear signal with small backgrounds. There are 10,471$\pm$121 events in the 1~fb$^{-1}$ sample. The signal lifetime distribution along with the backgrounds and the lifetime fit are shown in Fig.~\ref{KK-life}(right). The lifetime acceptance is determined by using a procedure called ``swimming" of the primary vertex, developed by previous experiments \cite{Bailey:1985zz,*Adam:1995mb,*Aaltonen:2010ta}.  For each decay a per-event acceptance function is determined by moving the primary vertex along the momentum vector of the $B$ particle. 
The measured lifetime is $1.407\pm 0.016\pm 0.007$\,ps. LHCb uses their measurements of $\Gamma_s$ and $\Delta\Gamma_s$ and an approximate first order equation, $\Gamma[f,t] \propto e^{-\Gamma_s t}\left[1+{\cal{A}}_{\Delta\Gamma}\left(\frac{\Delta\Gamma_s}{2}t\right)\right]$, to derive a value of ${\cal{A}}_{\Delta\Gamma}=-0.87\pm 0.17\pm 0.13$. A value of --1 is expected if there is no \CP 
violation.\footnote{Another mode that could have a significant Penguin amplitude contributions is $\Bsb\to\jpsi K_S^0$. The effective lifetime in this mode has been measured as $1.75\pm 0.12\pm 0.07$\,ps \cite{Aaij:2013eia}. Due to the relatively large uncertainty, this mode is not discussed further.}

\begin{figure}[h!]
\vspace{2mm}
\centering
\includegraphics[width=1.0\textwidth]{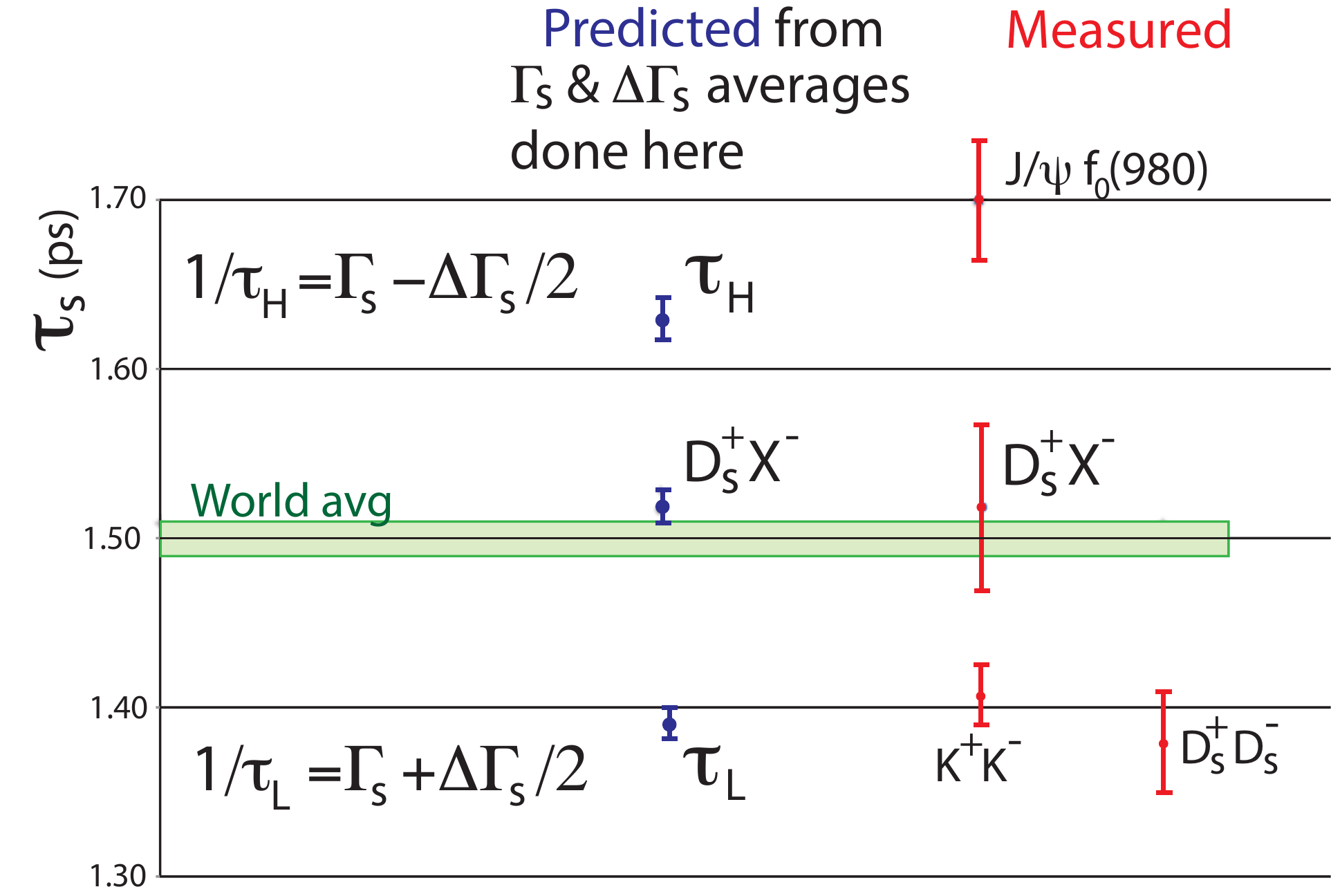}
\vspace{-6mm}
\caption{World average measurement of $\tau_s$ using the values from Table~\ref{tab:Gammas}. The middle column shows predictions for \Bsb lifetime values in \CP eigenstate and flavor specific modes (blue). The right column gives the measured values (red). The data values for the final states come from: $\jpsi f_0(980)$ \cite{Aaltonen:2011nk,Aaij:2012nta}, $D_s^{\pm}X^{\mp}$ \cite{Aaltonen:2011qsa}, $K^+K^-$ an LHCb preliminary result, and $D_s^+D_s^-$ \cite{Aaij:2013bvd}. }
\label{tau_s-other}
\end{figure}
From the measurements of $\Gamma_s$ and  $\Delta\Gamma_s$ discussed above in the $\jpsi K^+K^-$ and $\jpsi\pi^+\pi^-$ modes, the lifetimes of  $\Bsb$ \CP eigenstate and flavor specific modes can be predicted if Tree diagrams are dominant. Then the lifetimes in the \CP eigenstates are given by $1/\tau_{L \atop H}=\Gamma_s\pm \Delta\Gamma_s/2$. The predictions and measured results are shown in Fig.~\ref{tau_s-other}. For the \CP eigenstates, \CP violation is taken to be zero, consistent with the current measurements. For the flavor specific prediction Eq.~\ref{eq:fs} is used.

The measured values for the \CP eigenstate lifetimes are in good agreement with the``predicted" values. This is expected for all the modes except $K^+K^-$, which also is in good agreement showing that ${\cal{A}}_{\Delta\Gamma}$ is consistent with not being affected by \CP violation. I conclude that we now have a good experimental understanding of the \Bsb lifetime and width difference.

\section{The heavy quark expansion (HQE)}
The use of the optical theorem and the operator product expansion leads to a theoretical prediction for the decay width, and hence lifetime, for each $b$-flavored hadron \cite{Lenz:2014jha}
\begin{align}
\Gamma=&\frac{G_F^2m_b^5}{192\pi^3}\left|V_{cb}\right|^2\left\{c_{3,b}\left[1-\frac{\mu_{\pi}^2-\mu_{G}^2}{2m_b^2}+{\cal{O}}
\left(\frac{1}{m_b^3}\right)\right] \right. \nonumber\\
 &+\left. 2c_{5,b}\left[\frac{\mu^2_G}{m_b^2}+{\cal{O}}\left(\frac{1}{m_b^3}\right)\right]+\frac{c_{6,b}}{m_b^3}
 \frac{\bra{B}(\overline{b}q)_{\Gamma}(\overline{q}b)_{\Gamma}\ket{B}}{M_B}+...
 \right\}.
\end{align}
Note first of all that there are no ${\cal{O}}\left(\frac{1}{m_b}\right)$ correction terms. The $c_{i,b}$ coefficients, except for $c_{3,b}$, and the matrix elements are different for each $b$-hadron. (The terms $\mu_{\pi}$ and $\mu_{G}$ are matrix elements of the kinetic operator and chromomagnetic operator, respectively.) This theory is used to extract values of $|V_{ub}|$ and $|V_{cb}|$ from inclusive $B^-$ and $\Bzb$ semileptonic decays, so its verification is of prime importance. 

The HQE was first invented circa 1986 \cite{Shifman:1984wx,*Bigi:1991ir,*Bigi:1992su}. Predictions were made using available calculations for the $c_{i,b}$ and matrix elements. One such set of predictions was \cite{Shifman:1986mx} that $\tau(\Bsb)/\tau(\Bzb)\approx 1.0$, $\tau(\Bm)/\tau(\Bzb)\approx 1.1$ and $\tau(\Lb)/\tau({\Bzb})\approx 0.96$.

\section{\boldmath The saga of the \Lb lifetime}

Tests of the HQE using lifetime measurements started in the 1990's.
The theory was improved by further calculations. For example, 
in the case of the ratio of lifetimes of the \Lb baryon, $\tau({\Lb})$, to the \Bzb meson, $\tau({\Bzb})$,
the corrections of order ${\cal{O}}(1/m^2_b)$ were found to be small,  initial estimates of ${\cal{O}}(1/m^3_b)$ \cite{Neubert:1996we,Uraltsev:1996ta,*DiPierro:1999tb}  effects were also small, thus differences of only a few percent were expected \cite{Neubert:1996we,Cheng:1997xba,*Rosner:1996fy}.
Measurements at LEP in the  indicated that $\tau({\Lb})/\tau({\Bzb})$ was significantly lower than the prediction: in 2003 one widely quoted average of all data
gave $0.798\pm0.052$ \cite{Battaglia:2003in}, while another gave  $0.786\pm0.034$ \cite{Tarantino:2003qw,*Franco:2002fc}.  Explanations of the small value of the ratio were attempted by including additional operators or other modifications \cite{Ito:1997qq,*Gabbiani:2003pq,*Gabbiani:2004tp}, while there was resistance by others, who thought that the
HQE could be pushed to provide a ratio of $\sim$0.9, but no smaller \cite{Uraltsev:2000qw}. 

More recent measurements showed indications that a higher value is possible \cite{Aad:2012bpa,*Chatrchyan:2013sxa,*Aaltonen:2009zn,*Aaltonen:2010pj,*Abazov:2012iy}, although the uncertainties of these measurements are large. The LHCb collaboration performed two measurements of the lifetime ratio using the $\Lb\to\jpsi pK^-$ decay, one using 1\invfb of data \cite{Aaij:2013oha} and the other using their 3\invfb sample \cite{Aaij:2014zyy}.\footnote{LHCb also made an absolute measurement of $\tau({\Lb})$ using the $\jpsi\Lz$ final state, but since the precision is much worse than the relative measurements, I will not discuss it \cite{Aaij:2014owa}.}  This \Lb decay mode was first seen by LHCb. It is quite useful for lifetime measurements as it contains four charged tracks from the $\Lb$ decay vertex the decay time resolution is a remarkable good 40\,fs. I will only discuss the 3\invfb measurement.

\begin{figure}[t!]
\begin{center}
    \includegraphics[width=0.9\textwidth]{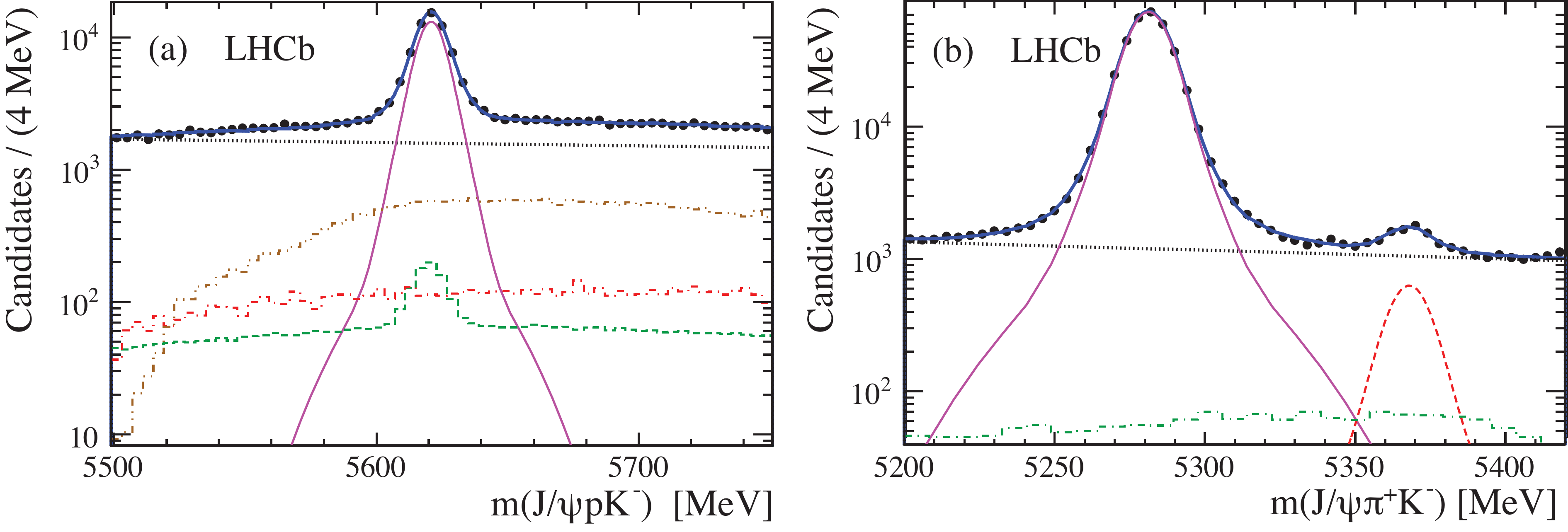}%
\end{center}
\label{fig:signal-log}
\vskip -0.5cm
\caption{ Fits to the invariant mass spectrum of  (a) $\jpsi pK^-$ and  (b)  $\jpsi \pi^+K^-$ combinations. The $\Lb$ and \Bzb signals are shown by the (magenta) solid  curves. The (black) dotted lines are the combinatorial backgrounds, and the (blue) solid curves show the totals. In (a) the $\Bsb \to \jpsi  K^+K^-$ and $\Bdb \to \jpsi \pi^+K^-$ reflections, caused by particle misidentification, are shown with the (brown) dot-dot-dashed and (red) dot-dashed  shapes, respectively, and the (green) dashed shape represents the  doubly misidentified $\jpsi K^{+}\overline{p}$ final state, where the kaon and proton masses are swapped.  In (b) the $\Bs\to \jpsi \pi^+K^-$ mode is shown by the (red) dashed  curve and the (green) dot-dashed shape represents the $\Lb\to\jpsi p\Km$ reflection.}
\end{figure}
 In this measurement the \Lb decay time distribution is compared to that of $\Bzb\to\jpsi K^-\pi^+$ decays. The reconstructed invariant mass distributions for both modes  are shown in Fig.~\ref{fig:signal-log}.  For \Bzb candidates the invariant $\pi^+ K^-$ mass was required to be within $\pm 100$ MeV of the $\Kstarzb(892)$ mass. There are approximately 50,000 \Lb signal events and 340,000 \Bzb signal events.
 
\begin{figure}[!t]
\vspace{-1cm}
\begin{center}
    \includegraphics[width=0.6\textwidth]{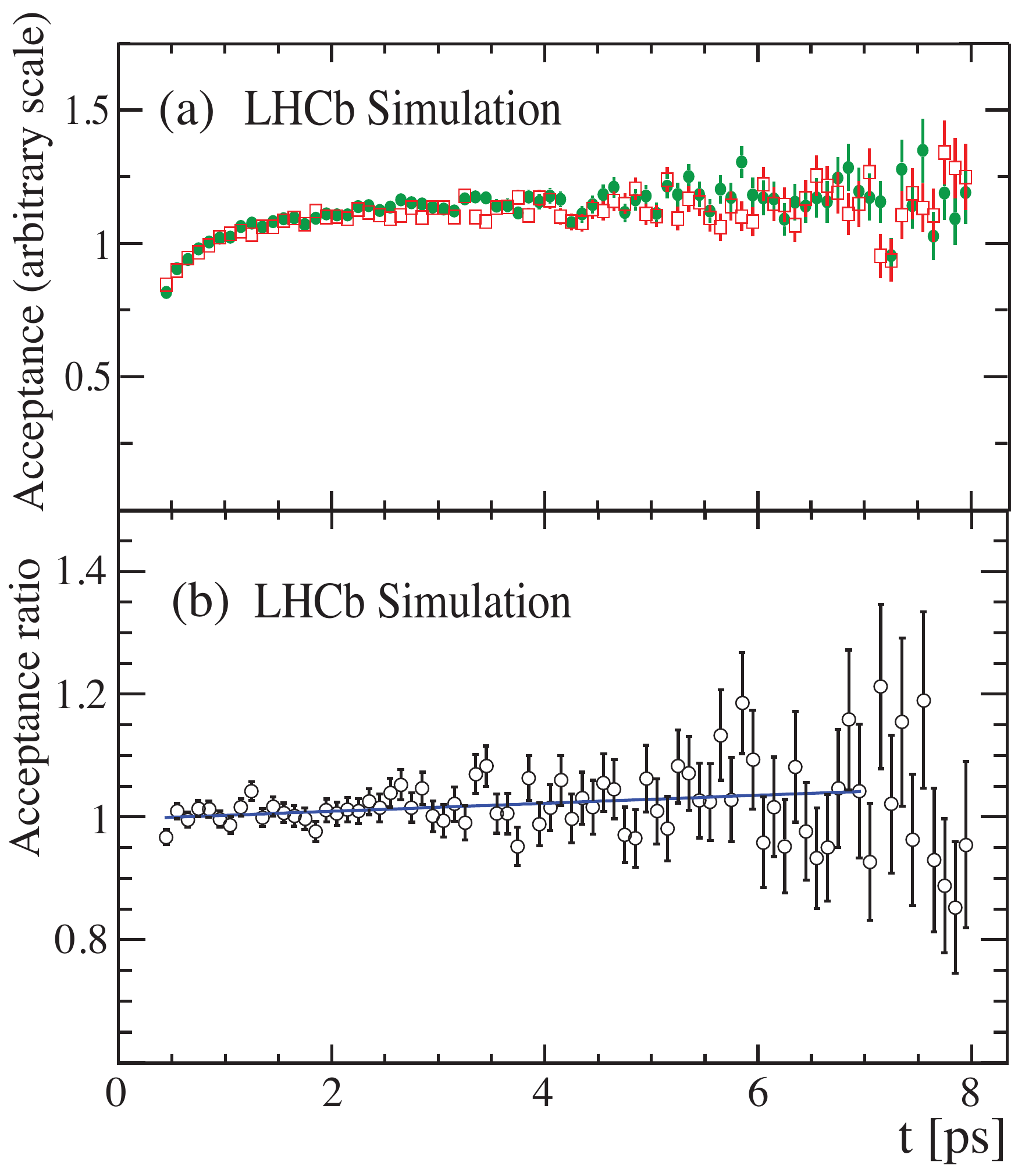}
\end{center}\label{fig:acceptances}
\vskip -0.5cm
\caption{ (a) Decay time acceptances (arbitrary scale) from simulation for (green) circles $\Lb \to \jpsi pK^-$, and  (red) open-boxes $\Bdb \to \jpsi \Kstarzb(892)$  decays. (b) Ratio of  the decay time acceptances between $\Lb \to \jpsi pK^-$ and  $\Bdb \to \jpsi \Kstarzb(892)$ decays obtained from simulation. The (blue) line shows the result of the linear fit.}
\end{figure}

The decay time acceptances obtained from the simulations are shown in Fig.~\ref{fig:acceptances}(a).  The individual acceptances in both cases exhibit the same behaviour of decreasing below 1\,ps.  
The ratio of the decay time acceptances is shown in Fig.~\ref{fig:acceptances}(b). The decay time range is chosen between 0.4--7\,ps because the acceptance is poorly determined for larger decay times, while for smaller ones the individual acceptances decrease quickly. 

\begin{figure}[b!]
\vspace{-1cm}
\begin{center}
    \includegraphics[width=0.62\textwidth]{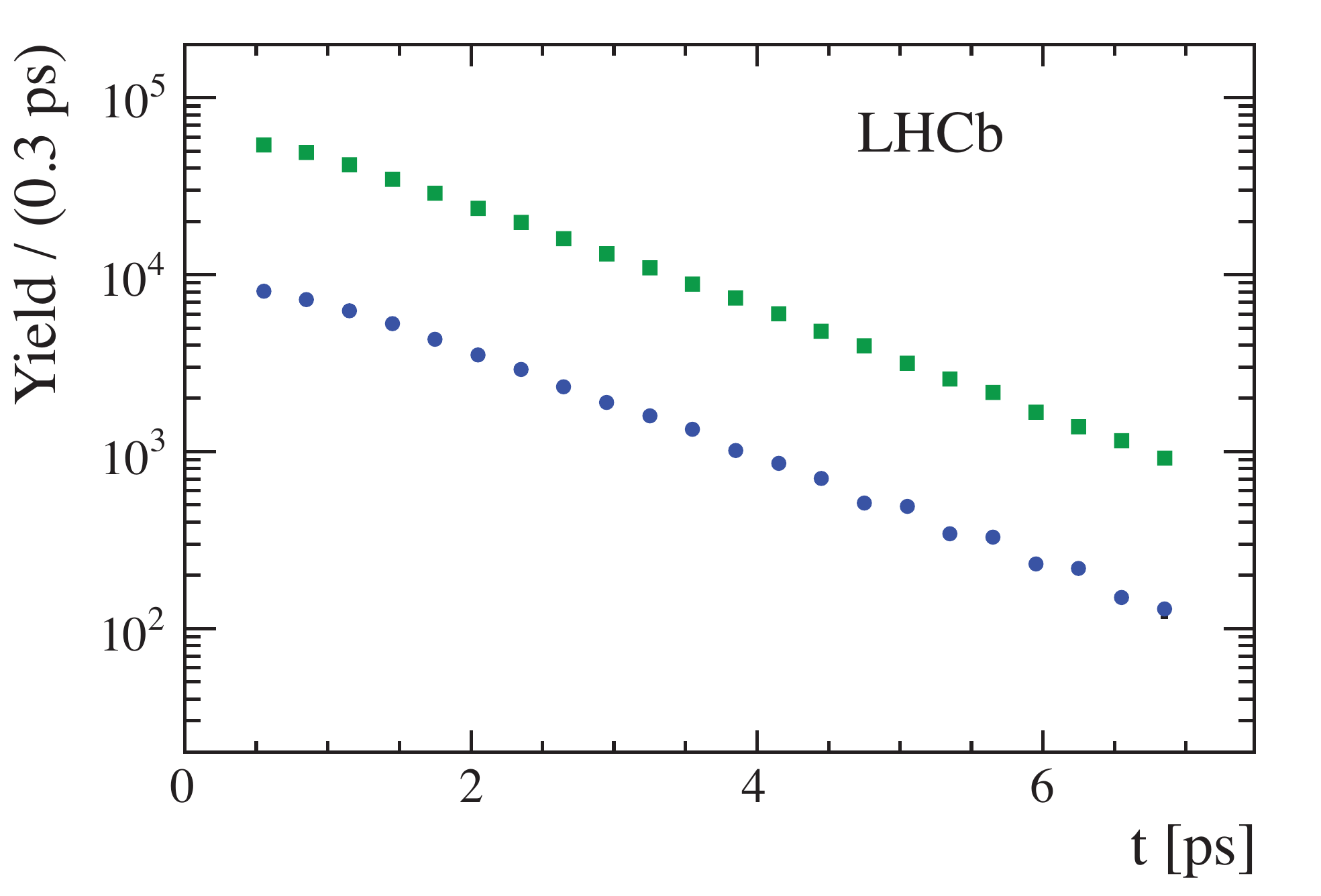}
\end{center}\label{fig:lifetime}
\vskip -.5cm
\caption{ Decay time distributions for  $\Lb \to \jpsi pK^-$ shown as (blue) circles, and   $\Bdb \to \jpsi \Kstarzb(892)$ shown as (green) squares. For most entries  the error bars are smaller than the points.}
\end{figure}

The yield of $b$ hadrons for both decay modes is determined by fitting the candidate invariant mass distributions in each decay time bin.
The resulting signal yields as a function of decay time are shown in Fig.~\ref{fig:lifetime}.

\begin{figure}[t!]
\begin{center}
    \includegraphics[width=0.9\textwidth]{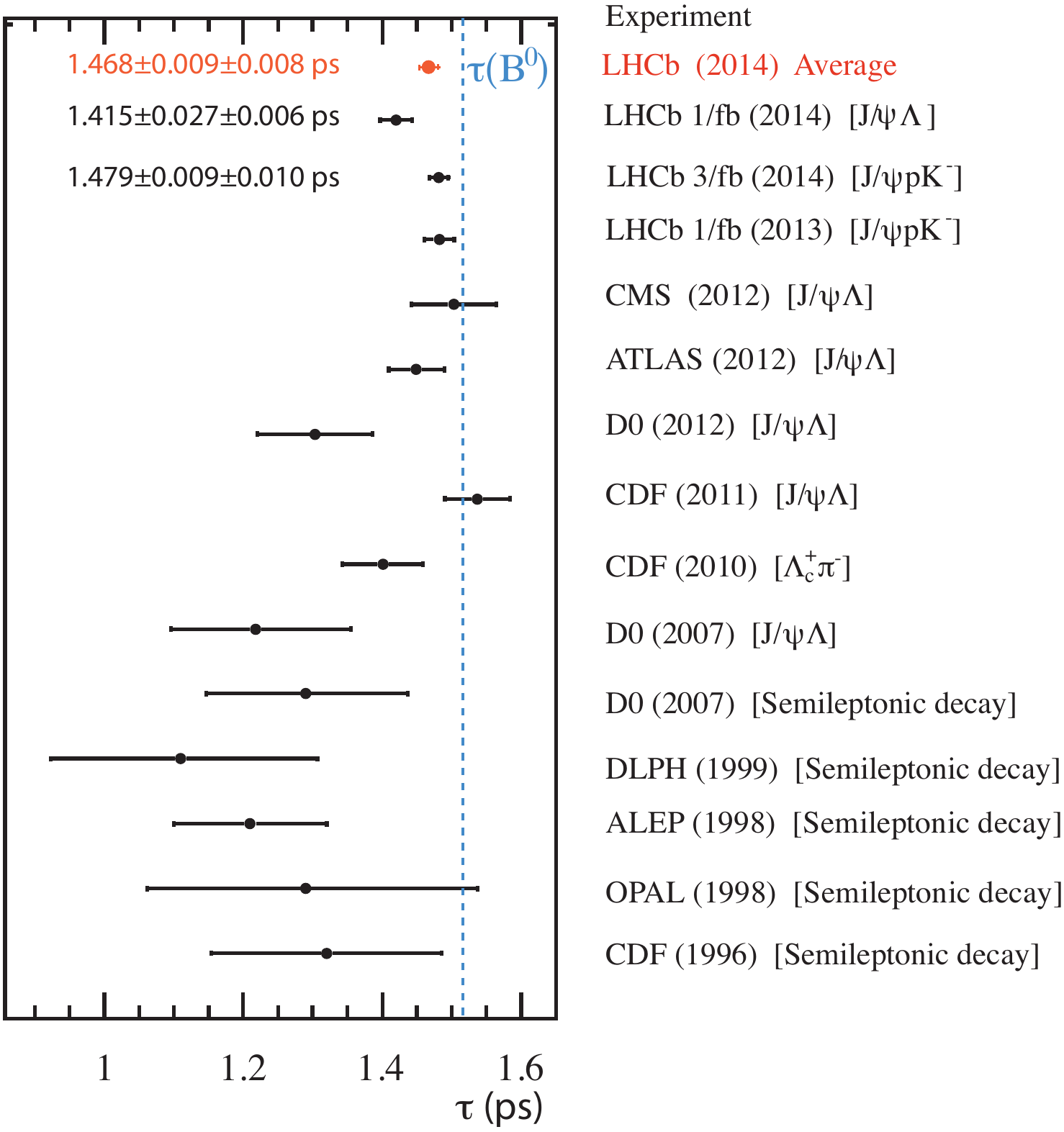}
\end{center}\label{PDG_lifetime-rev}
\vskip -0.5cm
\caption{Summary of measured \Lb lifetimes. The vertical dashed line shows the world average \Bzb lifetime.}
\end{figure}

The ratio of lifetimes is determined as
$\frac{\tau(\Lb)}{\tau(\Bdb)}=0.974\pm0.006\pm0.004$.
Multiplying the lifetime ratio by  $\tau(\Bdb)=1.519\pm 0.007$\,ps, the $\Lb$ baryon lifetime is 
$\tau(\Lb)= 1.479 \pm 0.009 \pm 0.010$\,ps.
A summary of \Lb lifetime measurements done since 1990 is shown in Fig.~\ref{PDG_lifetime-rev}. The LHCb recent values lead to the weighted LHCb average of
\begin{equation}
\tau(\Lb)= 1.468 \pm 0.009 \pm 0.008~{\rm ps}.
\end{equation}
In what follows I will use this value rather than an average of all the measurements as it is somewhat ad hoc to know which to include. It might be a good idea, however, to use only full reconstructed hadronic decays in averages whenever possible. The decay times in semileptonic decays must be corrected for missing particles and this processes could be biased and is subjected to additional systematic uncertainties.
\newpage
\section{\boldmath Measurement of the $\Xi^-_b$, $\Omega_b^-$ and $\Xi_b^0$ lifetimes.}

\begin{wrapfigure}{r}{3.0in}
\vspace{-5mm}
\centering
\includegraphics[width=0.2\textwidth] {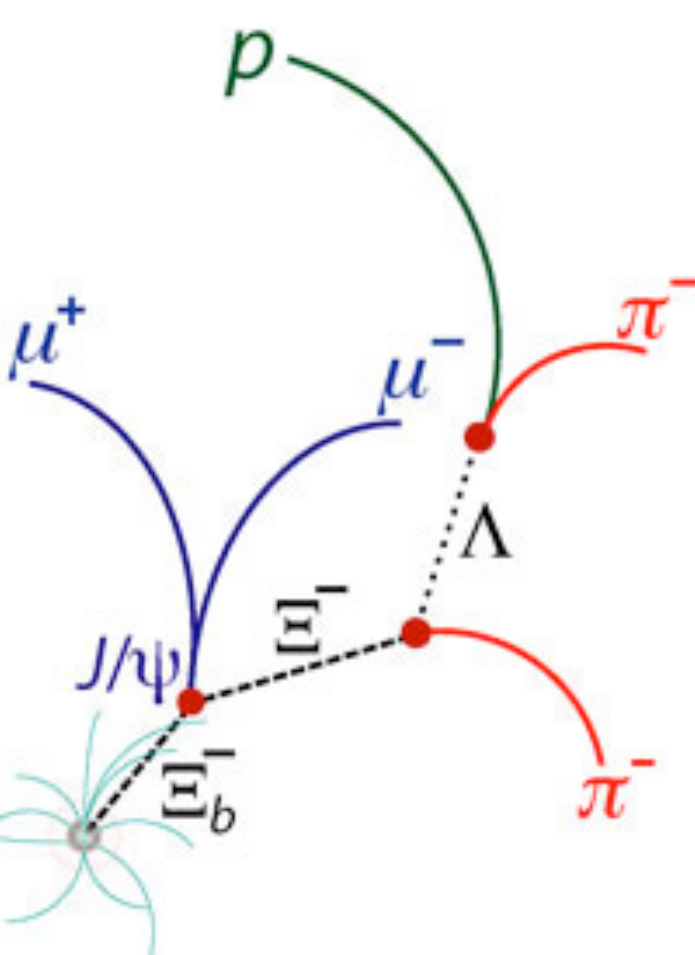}%
\vspace{-3mm}
\caption{Pictorial representation of $\Xi_b^-$ decay used in the CDF and LHCb lifetime measurements. For $\Omega_b^-$ the \pim is replaced with a $K^-$ and an $\Omega^-$ is reconstructed in place of a $\Xi^-$.}
\label{Xib}
\end{wrapfigure}
\begin{figure}[t!]
\centering
\includegraphics[width=0.45\textwidth] {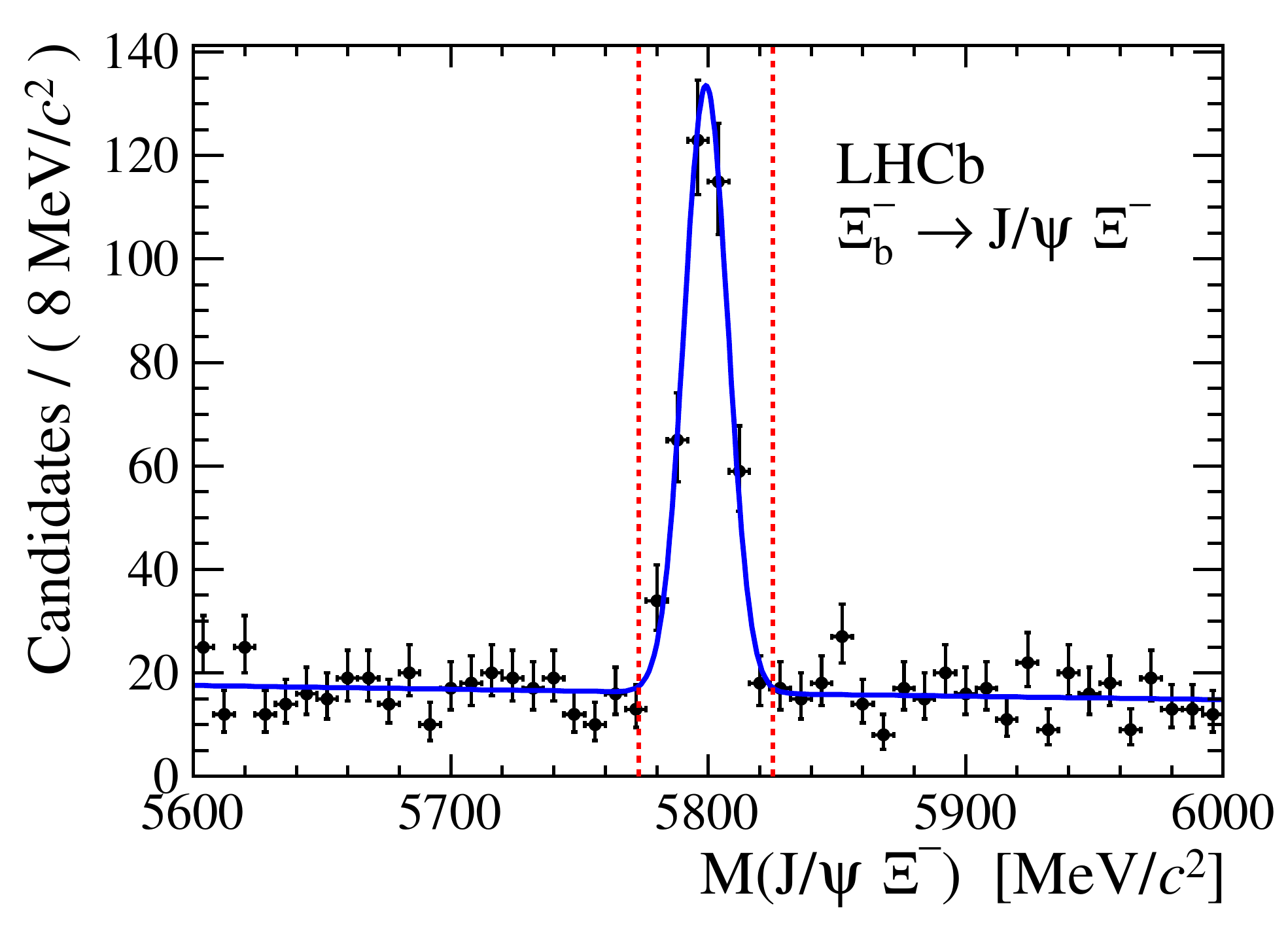}%
\includegraphics[width=0.45\textwidth] {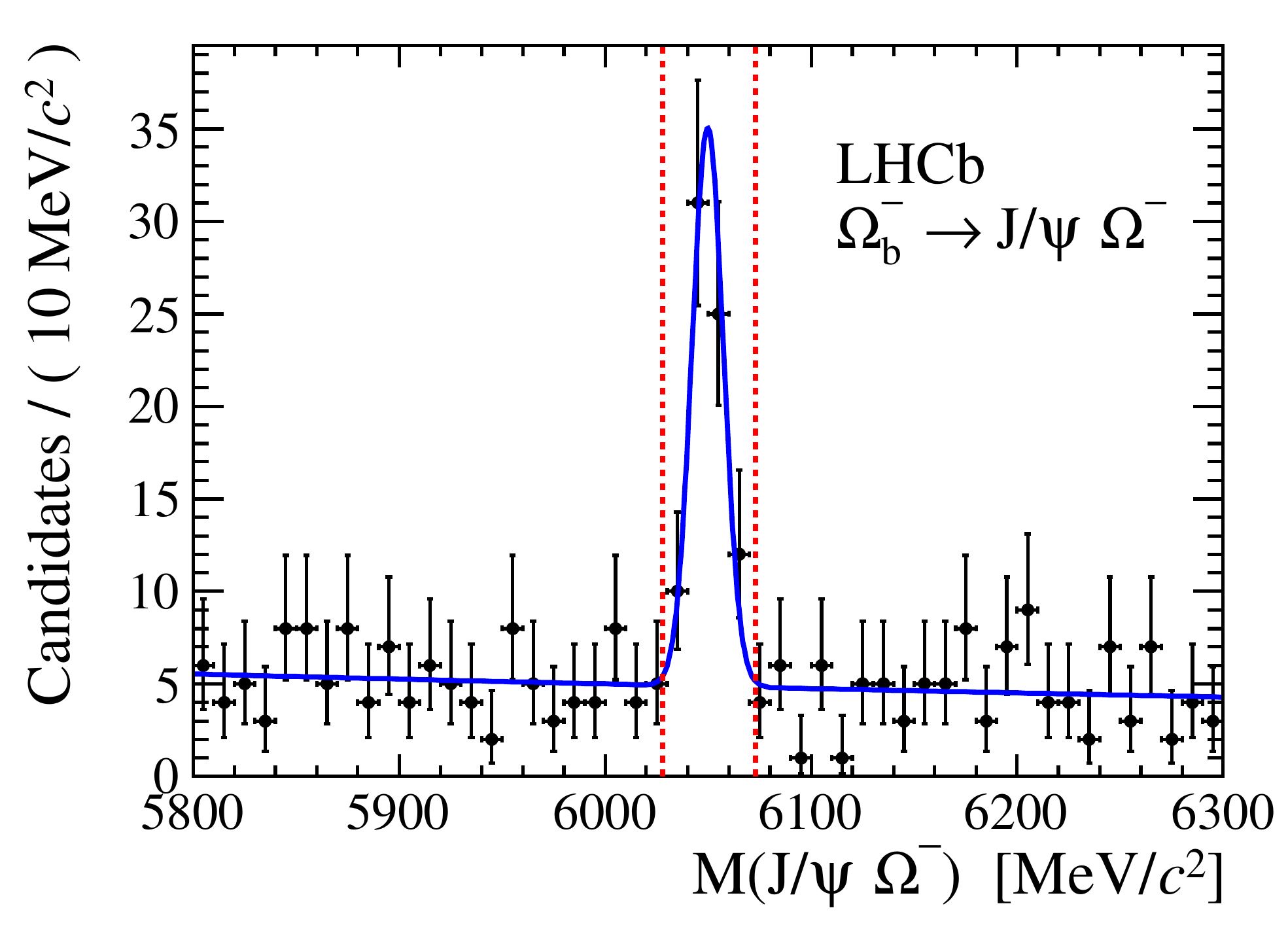}
\includegraphics[width=0.45\textwidth] {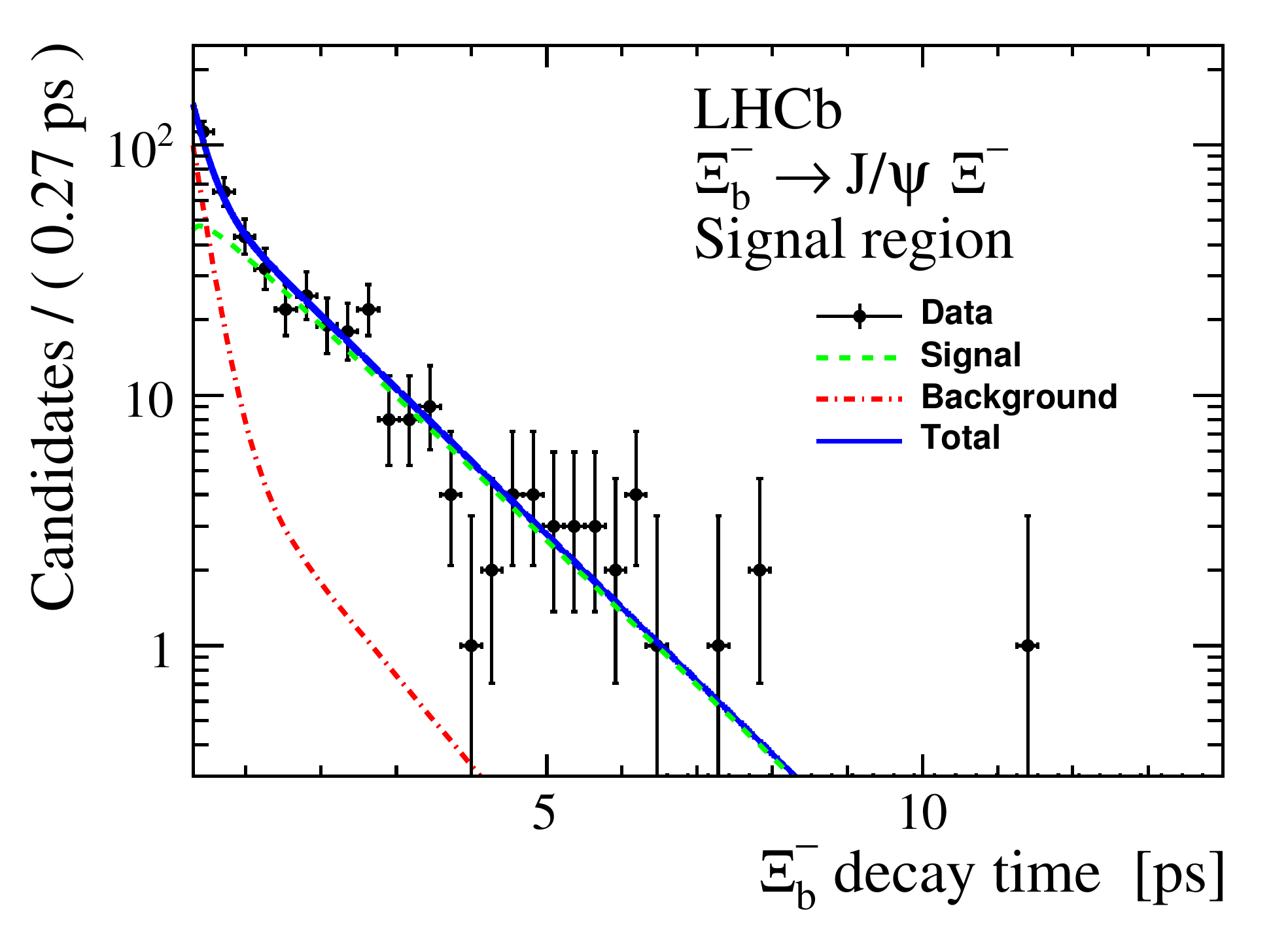}%
\includegraphics[width=0.45\textwidth] {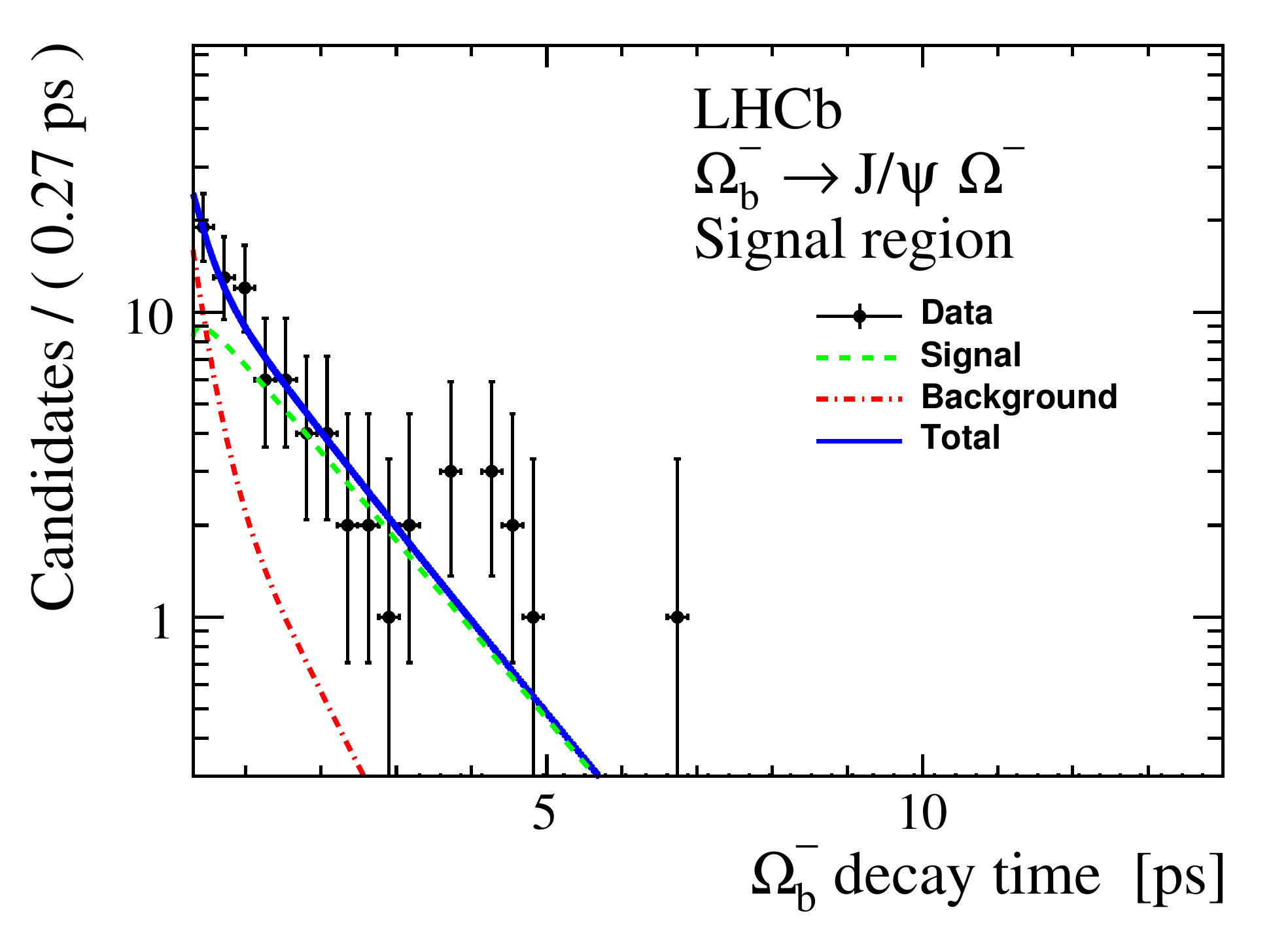}
\caption{Distributions of the reconstructed invariant mass (top) and decay time  of the $\Xi_b^-\to\jpsi \Xi^-$ (left) and $\Omega^-_b\to\jpsi\Omega^-$ (right) candidates. The signal mass region is defined as 5773--5825\, MeV for $\Xi_b^-$ and 6028--6073\,MeV for $\Omega^-_b$ candidates, shown by the vertical dotted lines. The results of the fits are overlaid.}
\label{XiO}
\end{figure}

The charged final states are found in the decay modes $\Xi^-_b\to \jpsi \Xi^-$, $\Xi^-\to \Lambda\pi^-$ and $\Omega^-_b\to \jpsi \Omega^-$, $\Omega^-\to \Lambda K^-$.  A pictorial diagram of the decay topology is shown in Fig.~\ref{Xib} for the $\Xi^-_b$ decay.
In both cases the $\Lambda$ is detected in the $p\pi^-$ decay mode. Absolute lifetime measurements have been made by CDF and LHCb. The LHCb $b$-baryon candidate invariant mass spectra and decay time distributions are shown in Fig.~\ref{XiO}. There are 313$\pm$20 $\Xi^-_b$ events, and 58$\pm$8 $\Omega_b^-$ events. The lifetime results are summarized in Table~\ref{tab:XiO}.

\begin{table}[htb]
\begin{center}
\begin{tabular}{l|cc}  
Exp. &  $\Xi^-_b$ lifetime (ps)&  $\Omega^-_b$ lifetime (ps)\\ \hline
CDF \cite{Aaltonen:2014wfa} & $1.32\pm 0.14\pm 0.02$& $1.66^{+0.53}_{-0.40}\pm 0.02$  \\
LHCb \cite{Aaij:2014sia}& $1.55^{+0.10}_{-0.09}\pm 0.03$ & $1.54^{+0.26}_{-0.21}\pm 0.05$  \\\hline 
Average&  $1.47\pm 0.08$ & $1.57\pm 0.21$ \\ \hline
\end{tabular}
\caption{Measurements of the $\Xi^-_b$ and $\Omega_b^-$ lifetimes.}
\label{tab:XiO}
\end{center}
\end{table}

\begin{figure}[b!]
\centering
\includegraphics[width=0.4\textwidth]{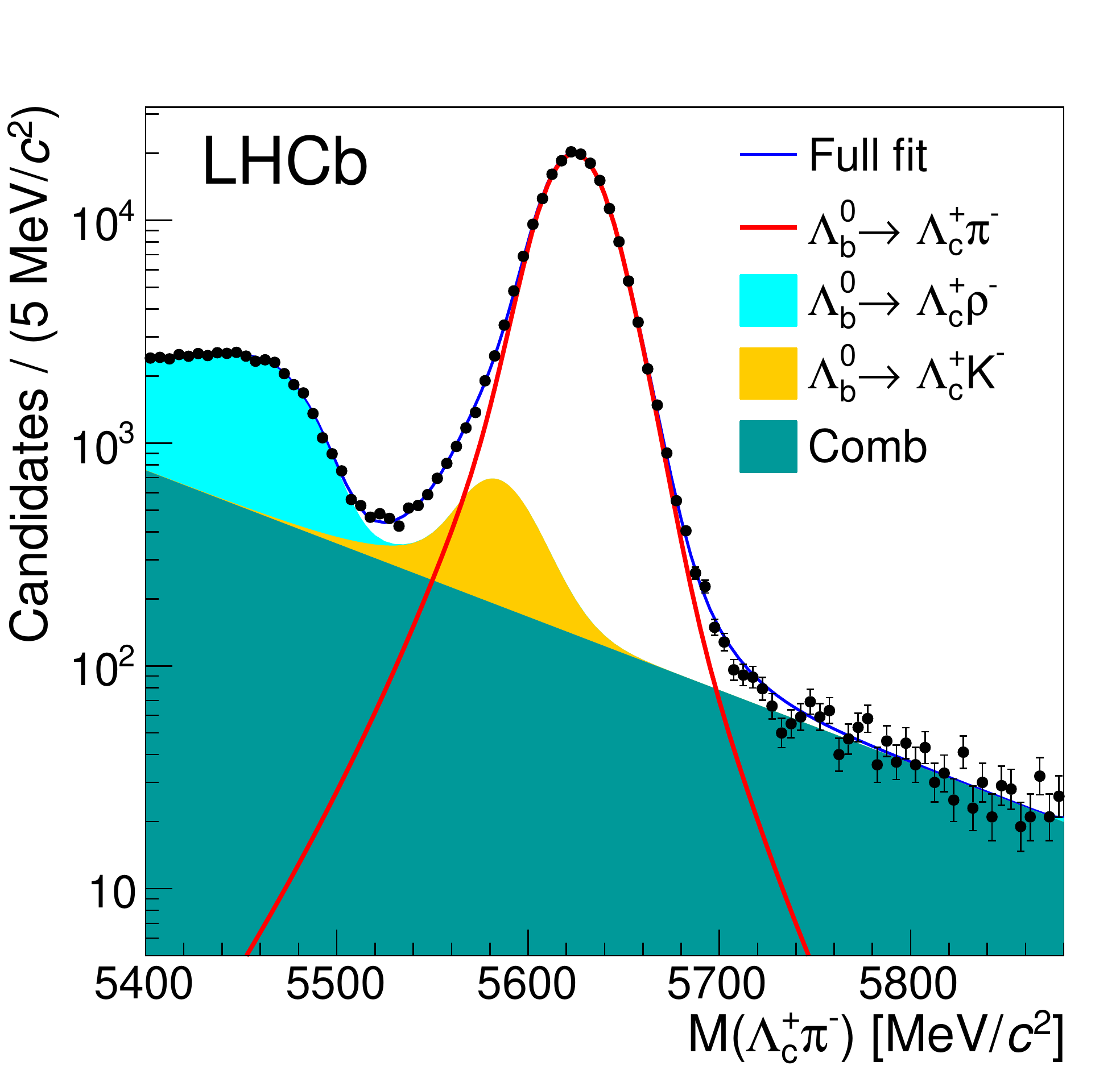}
\includegraphics[width=0.4\textwidth]{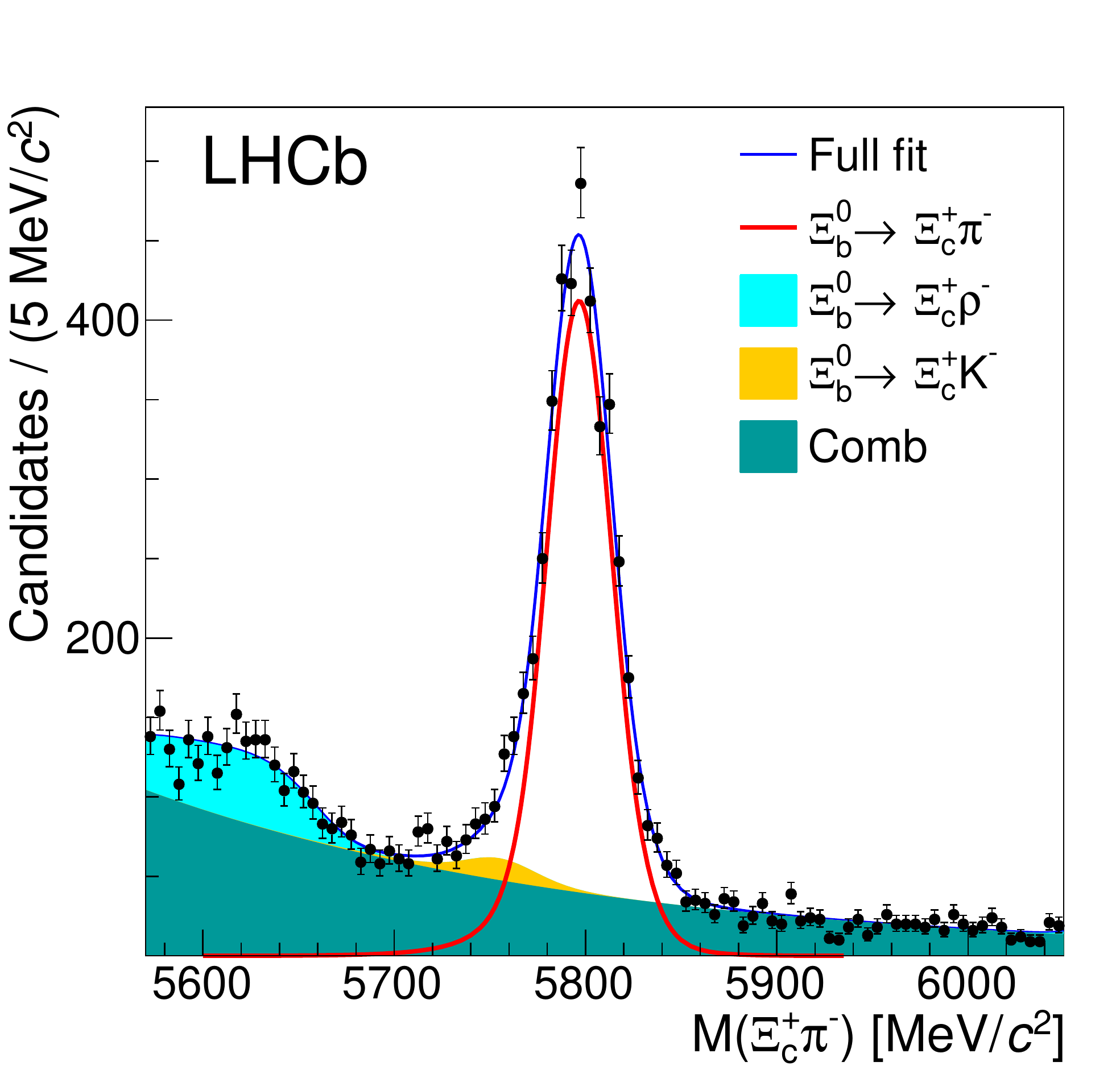}
\vspace{-2mm}
\caption{Invariant mass spectrum for (left) $\Lb\to\Lc\pim$ and (right) $\Xi^0_b\to\Xi_c^+\pim$
candidates along with the projections of the fit.}
\label{fig:XbMassFits}
\end{figure}

The relative lifetime of the $\Xi_b^0$ with respect to the \Lb has been measured by LHCb cleverly using two decay modes with exactly the same particle content namely, $\Xi^0_b\to \Xi_c^+\pi^-$, and $\Lb\to\Lc\pi^-$, with both $\Xi_c^+$ and \Lc decaying into $pK^-\pip$. Thus, the relative acceptance ratio is the consequence of different masses and different charm baryon lifetimes. The invariant mass for candidate decays is shown in Fig.~\ref{fig:XbMassFits}.

\begin{figure}[tb]
\centering
\includegraphics[width=1.0\textwidth]{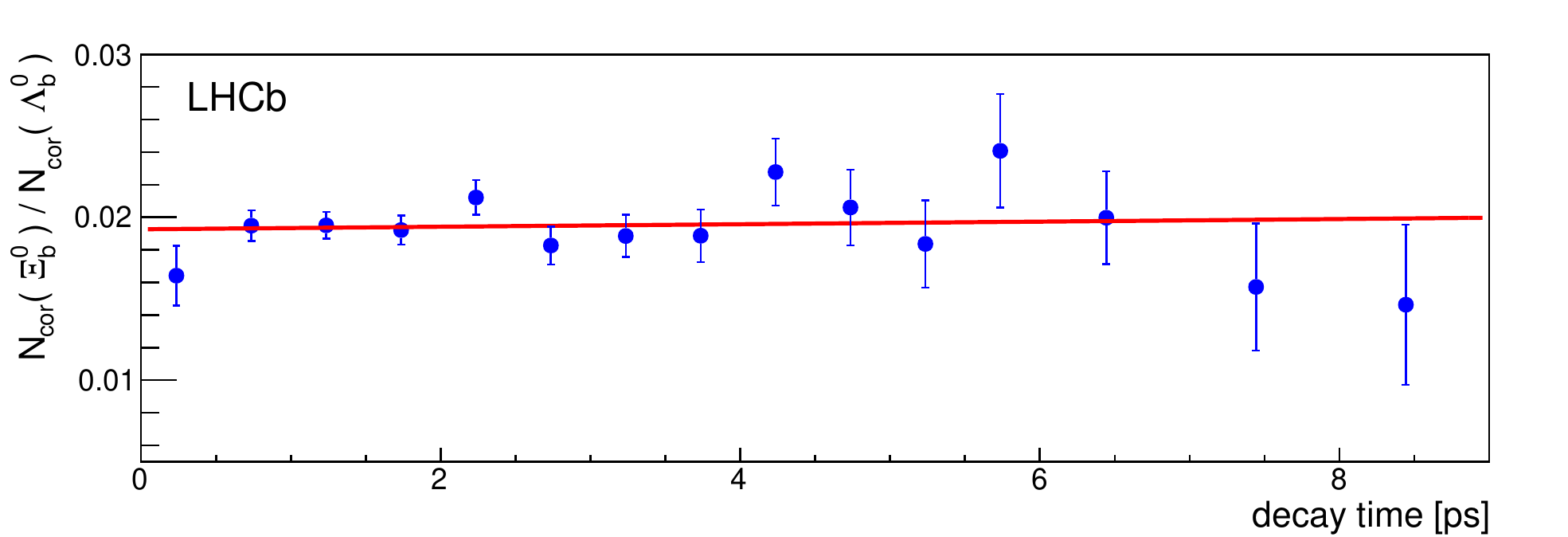}
\caption{\small{Efficiency-corrected yield ratio of $\Xi^0_b\to\Xi_c^+\pim$ relative to $\Lb\to\Lc\pim$ decays in bins of decay time. 
A fit using an exponential function is shown. The uncertainties are statistical only.}}
\label{fig:CorrYieldRatio}
\end{figure}

The measured yield ratio in each time bin is corrected by the relative efficiency of the two decay modes,
as obtained from simulated decays.
The efficiency-corrected yield ratio is shown in Fig.~\ref{fig:CorrYieldRatio}, along with the fit to an exponential function.
The points are placed at the weighted average time value within each bin, assuming an 
exponential distribution with lifetime equal to $\tau(\Lb)$. The bias due to this assumption is negligible.
From the fit, we find $\beta = (0.40\pm1.21)\times10^{-2}~{\rm ps}^{-1}$.
Using the measured $\Lb$ lifetime from  of 
$1.468\pm0.009\pm0.008$~ps, LHCb finds
\begin{align*}
\frac{\tau(\Xibz)}{\tau(\Lb)} &= 1.006\pm0.018\pm0.010,  \\
\tau(\Xibz) &= 1.477\pm0.026\pm0.014\pm0.013~{\rm ps},
\end{align*}
\noindent where the last uncertainty in $\tau(\Xibz)$ is due to the precision of
$\tau(\Lb)$.  

\section{Conclusions}

\begin{figure}[tb]
\centering
\includegraphics[width=1.0\textwidth]{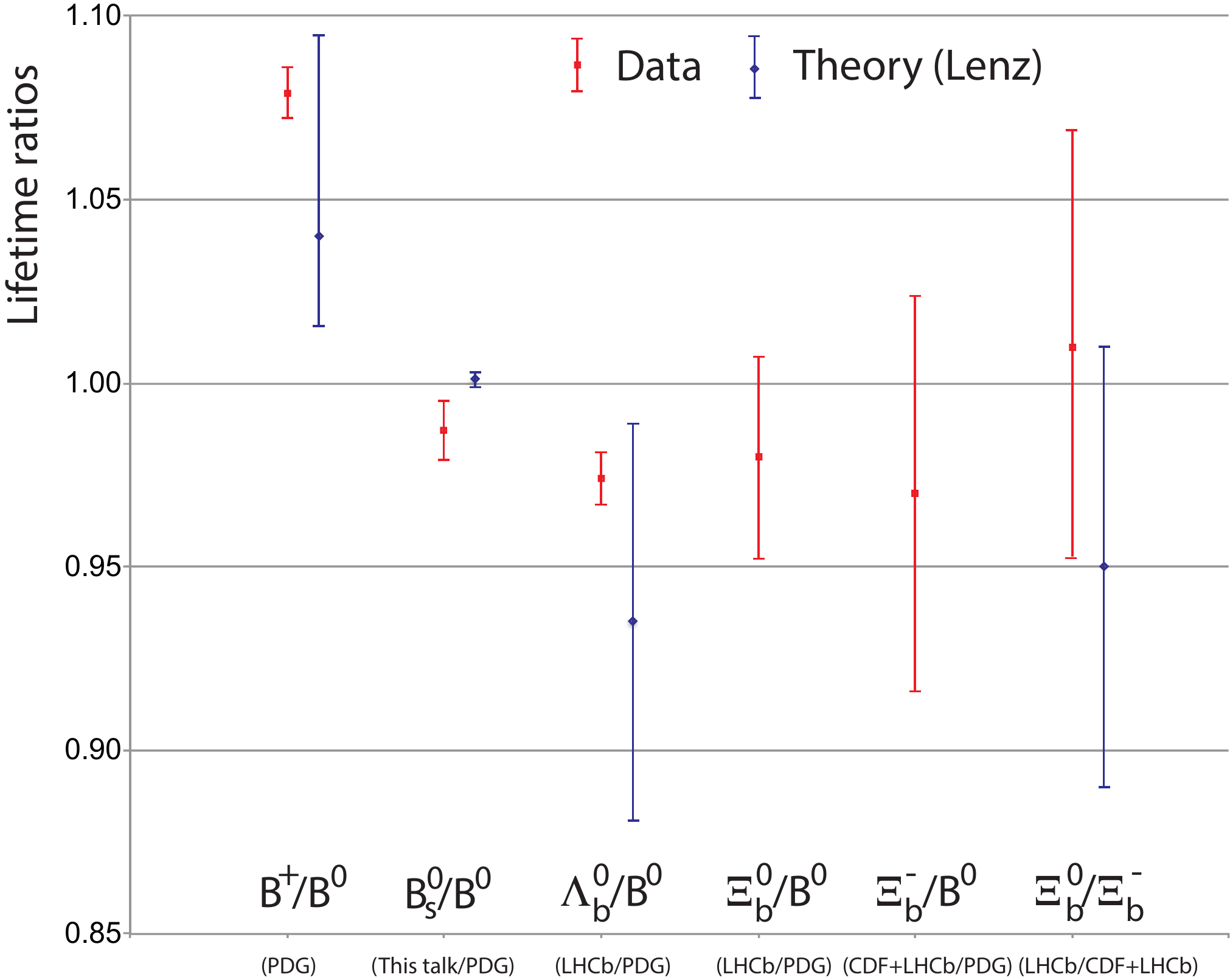}
\caption{Measured and predicted lifetime ratios using HQE \cite{Lenz:2014jha} for different $b$-flavored hadrons. }
\label{lifetime-ratios}
\end{figure}

Lifetimes from many new measurements of $b$-flavored hadrons have been presented. We have a good picture of the \Bsb lifetime including the value for $\Delta\Gamma_s$.\footnote{The value for $\tau({\Bsb})$ can be improved upon by judiciously averaging the values from the different decay modes that are not expected to be affected by Penguin contributions.} 
 The \Lb lifetime is now measured precisely and is consistent with the original predictions of the Heavy Quark Expansion, contrary to early indications. In Fig.~\ref{lifetime-ratios} I compare recent HQE predictions \cite{Lenz:2014jha} for the ratio of lifetimes with the measurements. The data is in good agreement with the theoretical predictions where they exist. Furthermore, except for the $\tau_{B^-}/\tau_{\Bzb}$ ratio, all the measurements with respect to $\tau_{\Bzb}$ are consistent with a value slightly smaller than one. The higher ratio for $\tau_{B^-}/\tau_{\Bzb}$ is explained using Pauli interference. The $\tau_{\Lb}/\tau_{\Bzb}$ ratio is measured much better than the theoretical prediction, so more effort here can be used to improve the calculation. For all the calculations, the current inaccuracy in theory is dominated
by the unknown non-perturbative matrix elements, which could be determined by lattice calculations.

In conclusion, a great deal of progress has been made recently understanding $b$-hadron lifetimes.

\afterpage{\clearpage}
\Acknowledgements
I thank the U.S. National Science Foundation for support. Useful conversations are acknowledged with Marina Artuso, Paolo Gandini, Alexander Lenz, and Olivier Leroy.

\newpage
\ifx\mcitethebibliography\mciteundefinedmacro
\PackageError{LHCb.bst}{mciteplus.sty has not been loaded}
{This bibstyle requires the use of the mciteplus package.}\fi
\providecommand{\href}[2]{#2}

\end{document}